\documentclass{article}

\usepackage{arxiv}

\usepackage[utf8]{inputenc} % allow utf-8 input
\usepackage[T1]{fontenc}    % use 8-bit T1 fonts
\usepackage[hidelinks]{hyperref}       % hyperlinks
\usepackage{url}            % simple URL typesetting
\usepackage{booktabs}       % professional-quality tables
\usepackage{nicefrac}       % compact symbols for 1/2, etc.
\usepackage{microtype}      % microtypography
\usepackage{graphicx}
\usepackage{doi}
\usepackage{upgreek}
\usepackage{multicol,multirow}
\usepackage{tabularx}
\usepackage{amsmath,amssymb,amsfonts}
\usepackage{mathrsfs}
\usepackage{amsthm}
\usepackage[figuresright]{rotating}
\usepackage[authoryear]{natbib}
\usepackage{newtxtext}
\usepackage{newtxmath}
\usepackage{textcomp}
\usepackage{xcolor}
\usepackage{graphicx, booktabs, multirow, makecell}
\usepackage{comment}
\usepackage{float}
\usepackage{subcaption}
\usepackage{placeins}
\usepackage{array}
\usepackage{pifont}
\usepackage{todonotes}
\usepackage{ulem}
\usepackage{makecell}

\title{Drag prediction of rough-wall turbulent flow using data-driven regression}

\date{} 					% Or removing it

\author{
\href{https://orcid.org/0000-0002-8542-523X}
{\includegraphics[scale=0.06]{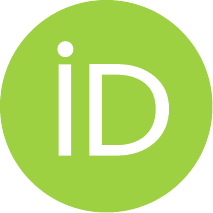}\hspace{1mm}Zhaoyu Shi}\\
	FLOW, Department of Engineering Mechanics\\
	KTH Royal Institute of Technology\\
	Stockholm 10044, Sweden \\
	\texttt{zhaoyus@kth.se} \\
	\And
	{\hspace{1mm}Seyed Morteza Habibi Khorasani} \\
	FLOW, Department of Engineering Mechanics\\
	KTH Royal Institute of Technology\\
	Stockholm 10044, Sweden \\
	\And
	{\hspace{1mm}Heesoo Shin} \\
	Mechanical Engineering Department\\
	Inha University \\
	Incheon 22212, Republic of Korea \\
	\And
	{\hspace{1mm}Jiasheng Yang} \\
	Institute of Fluid Mechanics\\
	Karlsruhe Institute of Technology\\
	Karlsruhe 76131, Germany \\
	\And
	{\hspace{1mm}}Sangseung Lee \\
	Mechanical Engineering Department\\
	Inha University \\
	Incheon 22212, Republic of Korea \\
        \And
        {\hspace{1mm}Shervin Bagheri}
        \thanks{corresponding author} \\
	FLOW, Department of Engineering Mechanics\\
	KTH Royal Institute of Technology\\
	Stockholm 10044, Sweden
}
\renewcommand{\shorttitle}{\textit{arXiv}}

\hypersetup{
pdftitle={A template for the arxiv style},
pdfsubject={q-bio.NC, q-bio.QM},
pdfauthor={David S.~Hippocampus, Elias D.~Striatum},
pdfkeywords={First keyword, Second keyword, More},
}

\begin{document}
\maketitle

\begin{abstract}
Efficient tools for predicting the drag of rough walls in turbulent flows would have a tremendous impact. However,  methods for drag prediction rely on experiments or numerical simulations which are costly and time-consuming. Data-driven regression methods have the potential to provide a prediction that is accurate and fast. We assess the performance and limitations of linear regression, kernel methods and neural networks for drag prediction using a  database of 1000 homogeneous rough surfaces. Model performance is evaluated using the roughness function obtained at friction-scaled Reynolds number 500. With two trainable parameters, the kernel method can fully account for nonlinear relations between $\Delta U^+$ and surface statistics (roughness height, effective slope, skewness, etc). In contrast, linear regression cannot account for nonlinear correlations and display large errors and  high uncertainty. Multilayer perceptron and convolutional neural networks demonstrate performance on par with the kernel method but have orders of magnitude more trainable parameters. For the current database size, the networks' capacity cannot be fully exploited, resulting in reduced generalizability and reliability. Our study provides insight into the appropriateness of different regression models for drag prediction. We also discuss the remaining steps before data-driven methods emerge as useful tools in applications. 
\end{abstract}

% keywords can be removed

\textbf{\mathversion{bold}Impact Statement} \par
The accurate estimation of drag in aviation and shipping is of great economic value as it significantly affects energy expenditure and carbon emissions. The long-standing pursuit of a universal correlation between drag and topographical features of roughness has received remarkable progress in recent decades, yet it is still limited by the feasibility of demanding experiments and simulations. There exists no model which is generally applicable to any given rough surface. This positions machine-learning (ML) modeling as a promising cost\nobreakdash-effective approach. Therefore, this study seeks to provide more insights into how different ML regression models perform in terms of the trade-off between capturing non-linearity and training costs. It is essential to scrutinize the correlation between the roughness features and drag before proceeding with the complete tuning and training of neural networks, as most relevant studies have overlooked. The comprehensive analysis presented in this work aims to offer valuable insights for the future design of ML-based models in the field of drag prediction.

\section{Introduction}\label{sec:intro}
Three-dimensional multi-scale surface irregularities are ubiquitous in industrial applications. The roughness imposes an increased resistance upon an overlying fluid flow, manifested as an increase in the measured drag. The increase in drag causes reduced energy efficiency, especially in turbulent flows. Examples include increased fuel consumption of cargo ships due to fouled hulls, reduced power output of eroded turbines in wind power plants, and an increase in the power input required to maintain a constant flow rate in pipelines with non-smooth walls. 

An efficient tool that can predict the drag induced by roughness would allow engineers and operators to optimize surface cleaning and treatment. However, as of today, there is no method for drag prediction that is efficient. The approaches that are in use today are accurate but also costly and time-consuming. They require towing tank experiments \citep{Schultz:2004} or direct numerical simulations  \citep{Thakkar:2016, Frooghi:2017, Thakkar:2018} of surface roughness replicas for extracting the equivalent sand grain roughness $k_s$. The extracted $k_s$ is used to estimate the drag penalty for simple geometries (e.g.~pipelines) or incorporated into computational fluid dynamics (CFD) software \citep{Bensow:2020, Marchis:2020} to evaluate the drag penalty on complex bodies (e.g. cargo ships).

Over time, a sufficient amount of relevant roughness data has accumulated, which can be used develop efficient regression models for predicting the drag of rough surfaces. Regression models can directly process images or topographical maps of the roughness to predict $k_s$, thus replacing experiments and resolved simulations. Recent efforts have focused on relatively complex regression methods. \cite{Jouybari:2021} adopted a multi-layer perceptron (MLP) and Gaussian processes regression to build a mapping from statistical surface measures to $k_s$. Both methods were trained on $45$ labelled samples, achieving an accuracy of approximately $10$\%. Realizing that the database size is the major bottleneck for fully exploiting the advantages of neural networks, \cite{Sangseung:2022} and \cite{Jiasheng:2023} employed transfer and active learning techniques, respectively. Specifically, \cite{Sangseung:2022} trained a MLP model on a small number of high-fidelity numerical simulations of synthetic irregularly rough surfaces to predict the roughness function, $\Delta U^+$. However, the model was pre-trained using estimates of drag obtained from empirical correlations for over $10000$ rough surfaces. \cite{Jiasheng:2023} used active learning, where the model automatically suggests the surface roughness that should be simulated and added to the database, to most effectively enhance the model performance. Previous studies have thus primarily investigated regression models based on neural networks, which --despite transfer and active learning-- require large datasets for reliable and accurate performance. It should be emphasized that drag prediction is a particularly demanding regression problem since each sample in the database used for training and testing is one DNS or experiment. Therefore, we are still far from having databases containing of the order of $10^4$ samples, which is commonly used for developing neural networks. 

Anticipating that data-driven models will eventually emerge as viable drag prediction tools, we assess the performance and limitations of increasingly complex regression methods. More specifically, this work compares linear regression, a kernel method based on support vector machine, multi-layer perceptron and a convolutional network. We have an order-of-magnitude larger database compared to earlier work \citep{Jouybari:2021, Sangseung:2022, Jiasheng:2023}. Using a GPU-accelerated numerical solver \citep{Costa:2021}, we developed a DNS database of $O(10^3)$ samples which includes five types of irregular homogeneous roughness. 

Alongside the actual technique used in regression, the choice of roughness features that constitute the model's input is another important aspect. For homogeneous roughness, the most common approach is to use statistics derived from the roughness height distribution, such as peak or peak-to-trough height \citep{Flack:2014, Frooghi:2017}, skewness \citep{Jelly:2018, Busse:2023} and effective slopes \citep{Jelly:2022}, etc. Given that rough surfaces in engineering applications often exhibit heterogeneous, e.g. patchy structures \citep{Busse:2023-2}, and anisotropy \citep{Frooghi:2017,Jelly:2022}, using the entire surface topography as input data may be needed to capture these complexities. In this paper, we will discuss different model inputs, including statistical measures and the two-dimensional height distribution of a surface.

This paper is organized into four sections: §\ref{sec:problem_set} describes the generation and statistical properties of the investigated database of rough surfaces. Model training and architecture details are outlined in §\ref{sec:models}. The drag prediction results of the modes are presented and discussed in §\ref{sec:discussion}. Finally, a discussion is provided in §\ref{sec:conclusion}.

\section{Problem setting} \label{sec:problem_set}
\subsection{Generation of irregular rough surfaces}
The dataset in this study includes five categories of irregular, statistically homogeneous rough surfaces. The surfaces are represented as a height function (or maps), $k(x,z)$, which is a function of streamwise ($x$) and spanwise ($z$) coordinates. Examples of topographies corresponding to each surface category are shown in the first column of table \ref{tab:surf_contour}, which displays representative height maps along with their projections onto $x-z$ plane. The rough surfaces of type $Sk_0$, are generated by specifying a Gaussian probability distribution of the roughness height \citep{Tevis:2017}. Accordingly, the mean skewness of all $Sk_0$-surfaces are approximately zero and the mean kurtosis is roughly three. By cutting off the heights below the average, we obtain the second type, i.e. the positively-skewed roughness ($Sk_+$) with mountainous topography. The negatively skewed surfaces ($Sk_-$) are generated in the opposite manner to $Sk_+$ and exhibit basins surrounded by flat regions. These three surface types are based on a prescribed skewness and are therefor \textit{isotropic}. Two other types of \textit{anisotropic} surfaces were generated using the MARS algorithm of \cite{Jelly:2018}. These are illustrated by the bottom two topographies in table \ref{tab:surf_contour}, which have streamwise- and spanwise-preferential effective slopes and are labelled as $\lambda_x$ and $\lambda_z$, respectively.
\begin{table}[t]
\centering
    \begin{tabular}{ccccc}
        \hline 
        \textbf{Exemplary Topography}  & \textbf{Type} & \textbf{No.} & \textbf{Properties} & \textbf{Isotropic}  \\
        \toprule
        \raisebox{-0.5\height}{\includegraphics[scale=0.2]{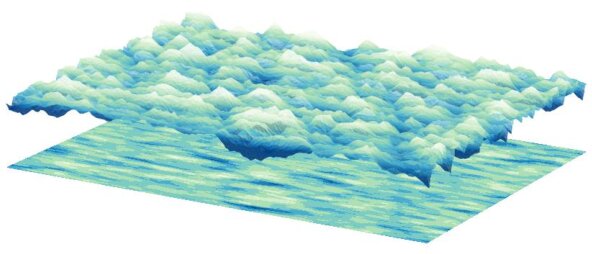}} & \makecell{$Sk_0$} & \makecell{271} & \makecell{$\langle Skw \rangle \approx -0.005$ \\ $\langle Ku \rangle \approx 2.98$ \\Gaussian} & \makecell{yes \\ $\langle ES_x/ES_z \rangle \approx 0.98$}   \\
        \midrule
        \raisebox{-0.5\height}{\includegraphics[scale=0.2]{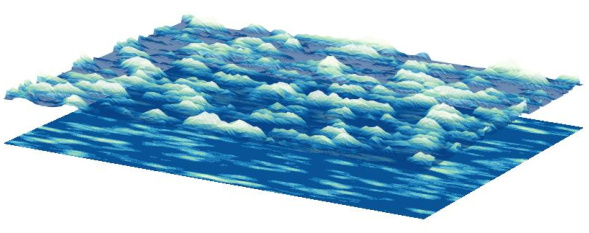}} & \makecell{$Sk_+$} & \makecell{200} &\makecell{$\langle Skw \rangle \approx 1.63$ \\ $\langle Ku \rangle \approx 5.34$ \\ Non-Gaussian} & \makecell{yes \\ $\langle ES_x/ES_z \rangle \approx 0.98$} \\
        % \cmidrule(r{6em}){1-2}
        \midrule
        \raisebox{-0.5\height}{\includegraphics[scale=0.2]{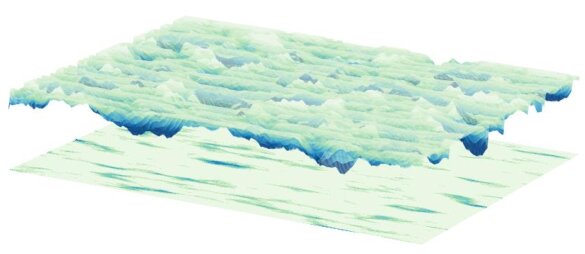}} & \makecell{$Sk_-$} & \makecell{141} & \makecell{$\langle Skw \rangle \approx -1.62$ \\ $\langle Ku \rangle \approx 5.23$ \\ Non-Gaussian} & \makecell{yes \\ $\langle ES_x/ES_z \rangle \approx 0.99$} \\
        \midrule
        \raisebox{-0.5\height}{\includegraphics[scale=0.2]{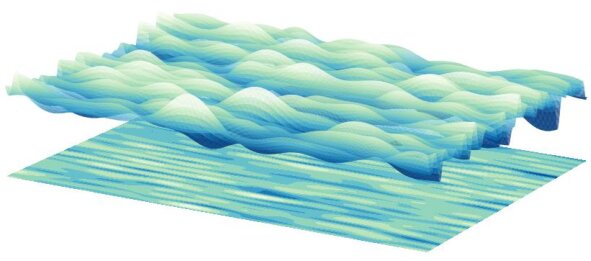}} & \makecell{$\lambda_x$} & \makecell{194} &\makecell{$\langle Skw \rangle \approx 0.009$\\ $\langle Ku \rangle \approx 2.98$ \\ Gaussian} & \makecell{no \\ $\langle ES_x/ES_z \rangle \approx 1.41$} \\
        \midrule
        \raisebox{-0.5\height}{\includegraphics[scale=0.2]{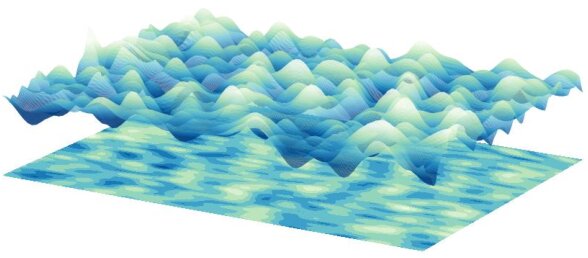}} & \makecell{$\lambda_z$} & \makecell{212} & \makecell{$\langle Skw \rangle \approx 0.005$ \\ $\langle Ku \rangle \approx 2.95$ \\ Gaussian} & \makecell{no \\ $\langle ES_x/ ES_z\rangle \approx 0.68$} \\
        \hline
    \end{tabular}
\vspace{1cm}
\caption{Examples of the five roughness types: the 3D topography of each type and their 2D projections on the $x-z$ plane are shown in the leftmost column. The number of samples of each type used in this study is given. The sample-averaged skewness and kurtosis $\langle \rangle$ are provided to demonstrate whether a surface is Gaussian or not in terms of its height distribution. Anisotropy is examined by the mean ratio of effective slopes over the samples in two directions.}
\label{tab:surf_contour}
\end{table}

We adopted a number of statistical measures for parameterizing the surface topographies as listed in table \ref{tab:stats}. The left panel of the table displays the seven parameters that characterize the topographical information of the surface, such as the effective slopes that represent the frontal solidity of the rough surfaces. \cite{Daniel:2021} provides a comprehensive summary of the physical significance of these statistical parameters. The center column displays three statistical measures of the topography's height distribution, of which the skewness has been shown to have a notable influence upon turbulent kinetic energy \citep{Thakkar:2016}, shear stress \citep{Jelly:2018}, and pressure drag \citep{Busse:2023}. Following \cite{Jouybari:2021}, we also use additional parameters formed from pairs of $ES_x$, $ES_z$, $Skw$, and $Ku$. These take into account nonlinear effects in the model input, the significance of which will be discussed later. 
\begin{table}[t]
\small
\centering
    \begin{tabular}{c|c|cc}
    \toprule
         \multicolumn{2}{c}{\textbf{Primary parameters}} & \multicolumn{2}{c}{\textbf{Pair parameters}} \\
         \cmidrule(rl){1-2}  \cmidrule(rl){3-4}
         \makecell{$k_c=k_{max}-k_{min}$  \\ $R_a=A^{-1}\int_{x,z}|k-k_{avg}|dA$ \\ $ES_x=A^{-1}\int_{x,z}|\frac{\partial k}{\partial x}|dA$ \\ $ES_z=A^{-1}\int_{x,z}|\frac{\partial k}{\partial z}|dA$ \\ $Po=(A\times k_c)^{-1}\int_{0}^{k_c}A_fdy$ \\ $inc_x=tan^{-1}\bigl\{\frac{1}{2}Skw(\frac{\partial k}{\partial x})\bigr\}$ \\ $inc_z=tan^{-1}\bigl\{\frac{1}{2}Skw(\frac{\partial k}{\partial z})\bigr\}$}
         & \makecell{$k_{rms}=\sqrt{\int_{x,z}(k-k_{avg})^2dA}$ \\
         $Skw=(A\times k_{rms}^3)^{-1}\int_{x,z}(k-k_{avg})^3dA$ \\ 
         $Ku=(A\times k_{rms}^4)^{-1}\int_{x,z}(k-k_{avg})^4dA$} 
         & \makecell{$ES_x^2$, $ES_z^2$, \\ $ES_x\cdot ES_z$, \\ $ES_x\cdot Skw$, \\ $ES_x\cdot Ku$, \\
        $ES_z\cdot Skw$, \\ $ES_z\cdot Ku$, \\ $Skw^2$, $Skw\cdot Ku$} \\
    \bottomrule
    \end{tabular}
\vspace{0.4cm}
\caption{The topographical statistics include ten `primary' parameters and nine `pair' parameters. The main features are divided into the ones bearing physical implications, i.e. crest height $k_c$, average height deviation $R_a$, effective slopes $ES_{x,z}$, porosity $Po$, inclinations $inc_{x,z}$; and statistical parameters, i.e. root-mean-square height $k_{rms}$, skewness $Skw$, and kurtosis $Ku$. }
\label{tab:stats}
\end{table}

\subsection{Drag measurement}
The drag penalty in turbulence from rough walls is commonly represented by the velocity deficit referred to as the roughness function $\Delta U^+=\Delta U/u_\tau$ \citep{Hama:1954}, i.e. the friction-scaled downward offset of the mean velocity profile in the logarithmic layer. Here, $u_\tau\equiv\sqrt{\tau_w/\rho}$ is the friction velocity, $\tau_w$ is the wall shear stress and $\rho$ is the fluid density.  

To determine the drag for each generated surface, direct numerical simulations (DNSs) of turbulent channel flow at $Re_\tau=u_\tau\delta/\nu=500$ were conducted (here, $\delta$ is half channel height). Considering the number of generated surfaces (1018), the simulations needed to be done in a cost-effective manner. For this reason, we employed the minimal\nobreakdash-span channel approach of \cite{Chung:2015} and \cite{Macdonald:2017}, which has proven to a successful method for characterizing the hydraulic resistance of rough surfaces under turbulent flow conditions. The minimal\nobreakdash-span approach exploits the fact that the flow retardation imposed by the roughness occurs close to it and this effect remains constant away from the roughness, manifesting as a downward shift in the logarithmic region of turbulent velocity profile, $\Delta U^+$, and otherwise known as the roughness function \citep{clauser:1954,Hama:1954}. Therefore, the measure of drag we acquired from the DNS was $\Delta U^+$. The size of the minimal channel in this work is $(L_x, 2\delta, L_z)=(2.4, 2, 0.8)$.
The simulations were conducted using the open-source code CaNS \citep{Costa:2021} which solves the incompressible Navier\nobreakdash-Stokes equations on three\nobreakdash-dimensional Cartesian grids using second\nobreakdash-order central finite\nobreakdash-differences. 

To incorporate the generated rough surfaces into the simulations, we augmented CaNS with the volume\nobreakdash-penalization immersed\nobreakdash-boundary method (IBM) \citep{Kajishima:2001,Breugem:2012}. Specifics regarding the solver and numerical methods used may be found in the aforementioned references which we omit here to avoid repetition. To ensure however that the DNS framework was able to accurately account for the effect of the irregular rough surfaces, a validation was carried out against one of the rough wall DNS cases of \cite{Jelly:2019}. The results of the validation are gathered in appendix \hyperref[app:validation]{\ref*{app:validation}}.
Details concerning the domain size and grid resolution that had to be satisfied when doing the minimal-span channel DNS of the 1000+ irregular rough surfaces can be found in the work of \cite{Sangseung:2022,Jiasheng:2022} and we refrain from stating them here.

\subsection{Parameter space}
The input to the different models is presented by $\mathbf{x}=(x_1,\dots, x_D)$, where $D$ is the number of input variables. For linear regression, SVR and MLP, we used the primary and secondary statistical measures in table \ref{tab:stats}, resulting in an input vector of size $D=10$ and $D=19$, respectively. For the CNN model, the roughness height map is used as the input, i.e. $D=n_zn_x=102\times 302$.
\begin{figure}
   \centering
   \includegraphics[width=.8\textwidth]{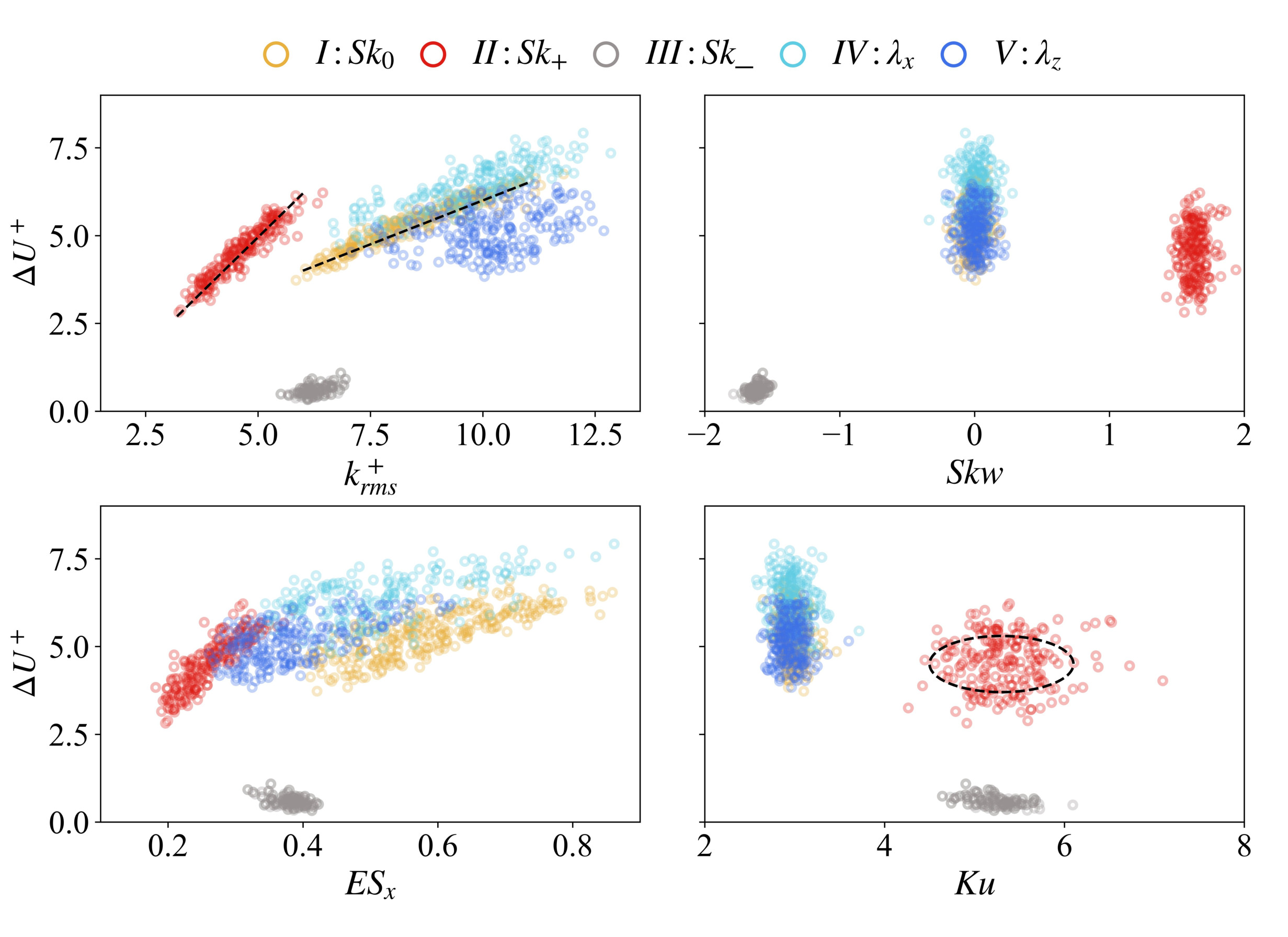}
   \put(-362, 240){(a)}
   \put(-185, 240){(b)}
   \put(-362, 120){(c)}
   \put(-185, 120){(d)}
   \caption{Scatter distributions of $\Delta U^+$ and four representative statistics of each type of roughness: (a) $k^+_{rms}$ (b) $Skw$ (c) $ES_x$ and (d) $Ku$. The dashed straight lines in (a) highlight the linear relationship between $\Delta U^+$ and $k^+_{rms}$ for merely zero- and positively-skewed surfaces while the 'cluster' distribution (circle lines) in (d) indicates a non\nobreakdash-linear relationship between $\Delta U^+$ and $Ku$ prediction.}
   \label{fig:param_scatter}
\end{figure}
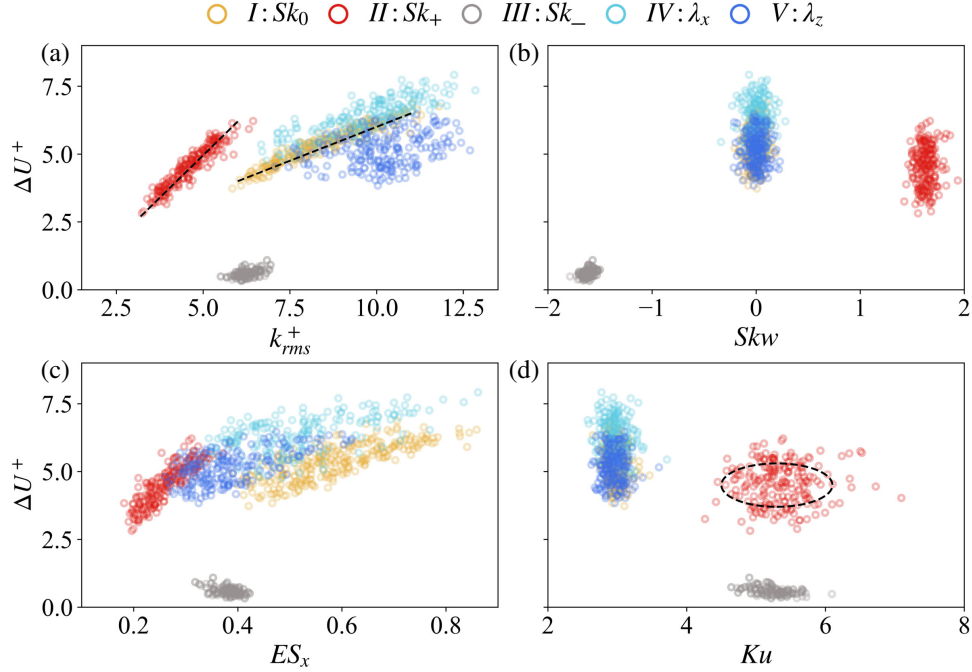

Before attempting any modeling for drag prediction, simply examining the distribution of input parameters with respect to the output provides insights into the relationship between them. Figure \ref{fig:param_scatter} shows the scatter distribution of four representative statistics ($k^+_{rms}, Skw, ES_x, Ku$) with $\Delta U^+$. A notable degree of linearity between $k^+_{rms}$ and $\Delta U^+$ exists for the $Sk_0$- and $Sk_+$-surfaces while it is less for the $\lambda_x$ and $\lambda_z$ surfaces (figure \ref{fig:param_scatter}a). Similarly, the effective slopes shown in figure \ref{fig:param_scatter}(c) show a certain degree of linearity with respect to $\Delta U^+$. However, the 'cluster' distributions seen for the skewness (figure \ref{fig:param_scatter}b) and kurtosis (figure \ref{fig:param_scatter}d) imply a nonlinear relationship that would need to be accounted for in any model for it to be more robust. Note that the negatively skewed roughness yields a much smaller $\Delta U^+$ compared to other types of roughness. These surfaces are dominated by pothole-like topography with few stagnation points. As a consequence, the viscous force contributes significantly to the total drag \citep{Busse:2023}, in contrast to the other surface types where pressure drag is dominant. Finally, we note from figure \ref{fig:param_scatter} that for our surfaces $\Delta U^+$ falls into the range of 0.1 to 7.5, thus including both transitionally and fully rough regimes \citep{Jimenez:2004}. By including this range of roughness, the models need to learn both viscous and pressure drag components. 

\begin{figure}
    \centering
    \includegraphics[width=0.85\textwidth]{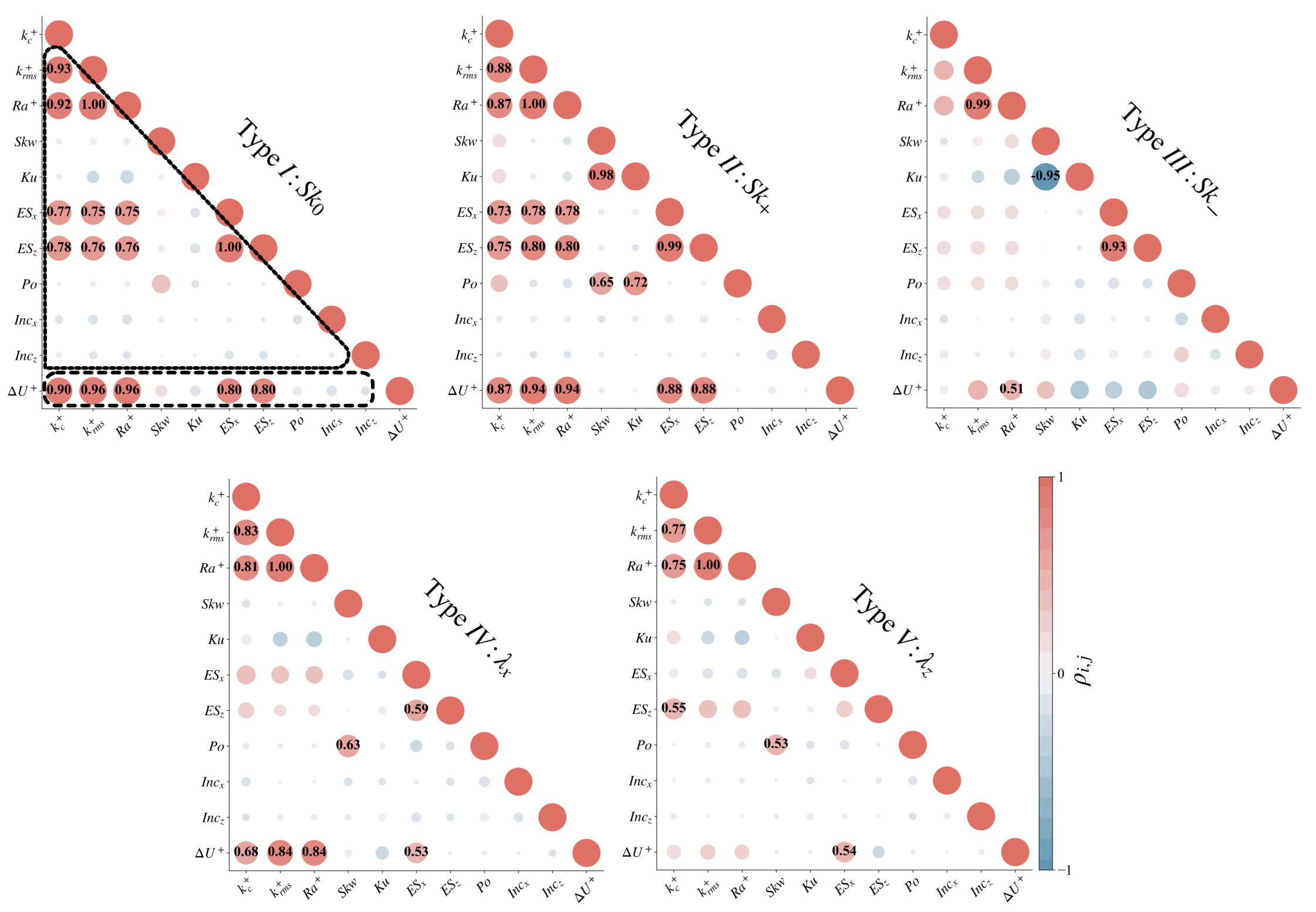}
    \put(-406, 266){(a)} \put(-273, 266){(b)} \put(-136, 266){(c)} 
    \put(-350, 125){(d)} \put(-220, 125){(e)} 
    \caption{Correlation coefficients $\rho$ of ten primary parameters and $\Delta U^+$ for each type of roughness.  The circles in the bottom row show the linear correlation between $\Delta U^+$ and the parameters while the rest are the correlations between any two topographical parameters. Larger and darker circles represent stronger linear correlation between two variables. Those with $|\rho_{ij}|>0.5$ are annotated.}
    \label{fig:corr_mat_map}
\end{figure}
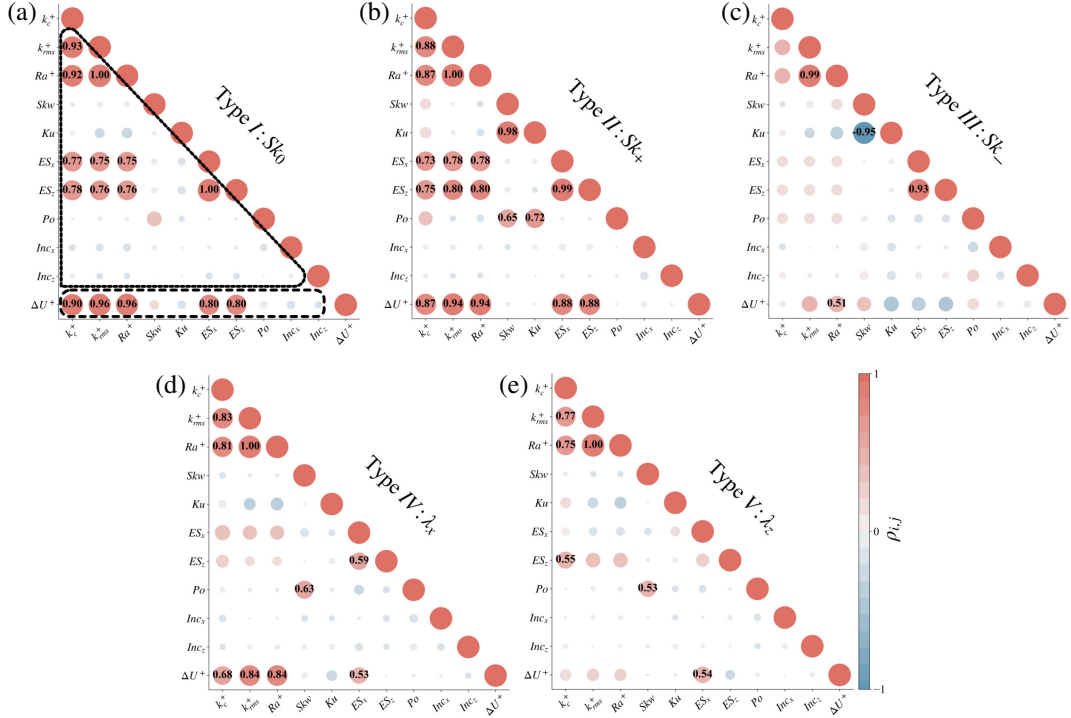

To further quantify the correlation between $\Delta U^+$ and the input parameters, we show in figure \ref{fig:corr_mat_map} the correlation coefficient $\rho(x_i, x_j)$, defined as
\begin{equation}
    \rho_{ij}=\frac{\sum(x_i-\langle x_i \rangle)(x_j-\langle x_j \rangle)}{\sqrt{\sum(x_i-\langle x_i \rangle)^2\sum(x_j-\langle x_j \rangle)^2}},
    \label{eq:pearson}
\end{equation}
where $x_i=k^+_c, k^+_{rms}, Skw,...,\Delta U^+$. The matrix in figure \ref{fig:corr_mat_map} visualizes the degree of linearitybetween the surface parameters (demarcated by the dashed triangle in figure \ref{fig:corr_mat_map}a) and also between the surface parameters and $\Delta U^+$ (bottom row in figure \ref{fig:corr_mat_map}a). Coefficient values greater than 0.7 indicate a strong linear correlation between two variables. The matrices of the $Sk_0$ and $Sk_+$ surfaces are overall similar, with the roughness height parameters ($k^+_c, k^+_{rms}, R^+_a$) and effective slopes ($ES_x, ES_z$) being strongly linearly correlated to $\Delta U^+$.
This is in contrast to other surface types which manifest a non-linear quality with respect to $\Delta U^+$. In particular, the $Sk_-$ surfaces exhibit a low degree of linearity both between parameters and parameter-$\Delta U^+$. This indicates a more intricate mapping between the surface properties of pitted surfaces and their resulting drag. For all the surface types, the roughness height and effective slopes are the parameters that exhibit a common degree of linear correlation to $\Delta U^+$. It is worth noting that, for the surfaces generated by the prescribed skewness (type I, IV, V), a weaker correlation between skewness and $\Delta U^+$ is observed. 
As illustrated in figure \ref{fig:param_scatter}(b), the range of $Skw$ for each type of surface is limited due to it being a prescribed (and hence controlled) parameter. This limited range precludes the possibility of revealing any relation between $Skw$ and $\Delta U^+$. This also applies to its correlation with other parameters, particularly for Gaussian surfaces.

\section{Predictive models} \label{sec:models}
\begin{figure}
    \centering
    \includegraphics[width=.75\textwidth]{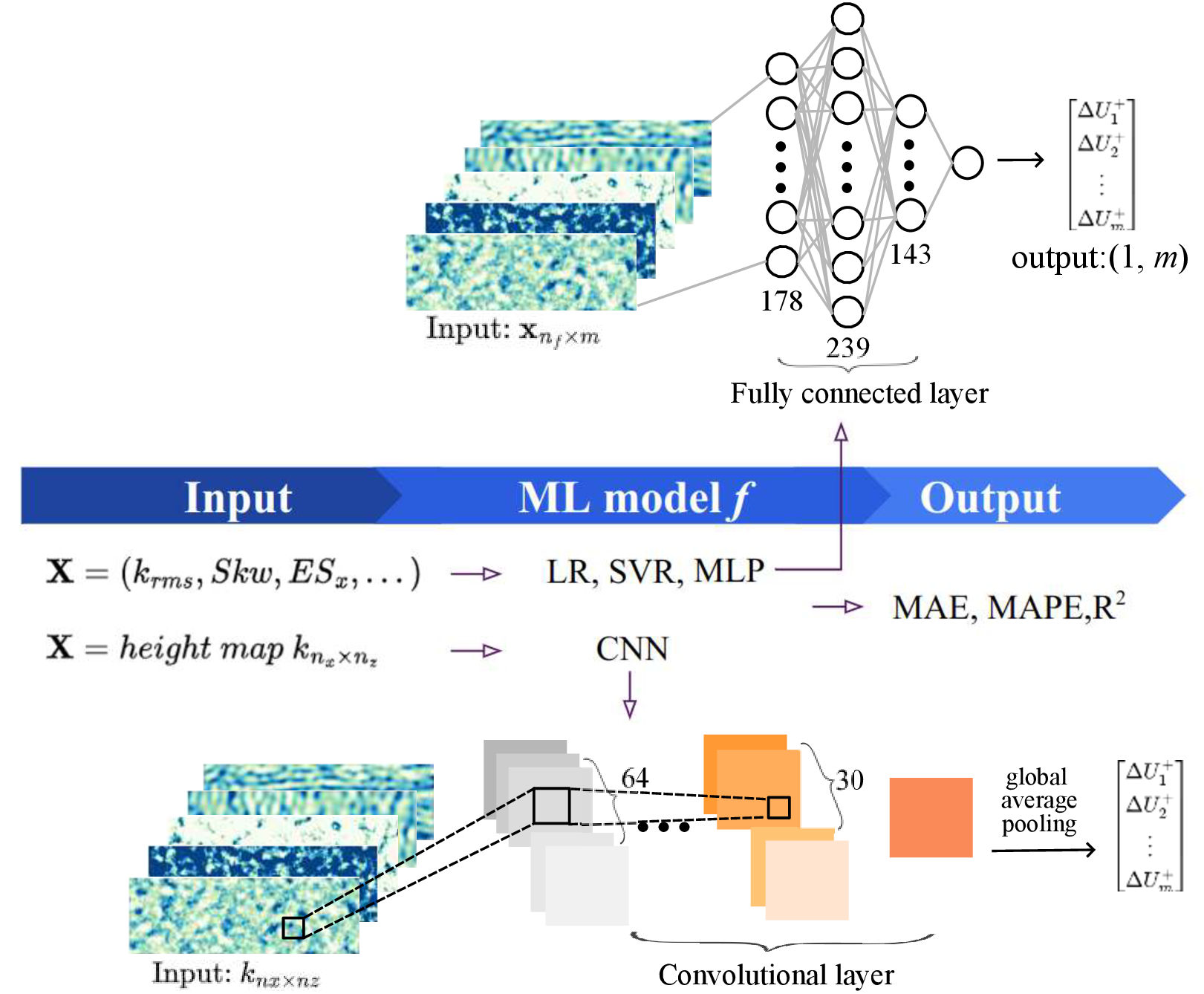}
    \caption{Workflow of drag prediction. The four models are evaluated by MAE, MAPE and $R^2$. The model architectures of MLP and CNN are illustrated, wherein the hyperparameters (HPs) are determined using Bayesian optimization. }
    \label{fig:flowchart}
\end{figure}

For training, we use a sequence of rough surfaces $\{\mathbf{x}_n\}$ together with their corresponding roughness functions $\{\Delta U^+_n\}$, where $n=1,\dots, N$ . The objective is to find the least complex model that accurately predicts the roughness function $\Delta \tilde U^+$ for a new (i.e. ``unseen'') rough surface, $\mathbf{x}$:
\begin{equation}
\Delta \tilde U^+=f(\mathbf{x}).
\label{eq:model}    
\end{equation}
Here, $f$: $\mathbb{R}^D\rightarrow\mathbb{R}$ represents different models obtained by solving a regression problem. We adopt the following approaches for creating the models: linear regression (LR), support vector regression (SVR) utilizing kernel functions, multi-layer perceptron (MLP), and convolutional neural network (CNN). Depending on the model, the inputs are either the statistical parameters listed in table \ref{tab:stats} (LR, SVR, MLP) or the height maps bearing the roughness topography (CNN).
We used 80\% of the total shuffled roughness data for training and validation with the remaining 20\% used for testing. A random sampling constituting 80\% of the development data is used for training, with each type of roughness comprising an equal fraction of this data. The data partitioning for training and testing is identical for all models. 

Figure \ref{fig:flowchart} illustrates the process of the regression modeling. For the neural networks, Bayesian optimization (BO) was used for hyperparameter tuning due to the large parameter space. The LR and SVR models were tuned manually. Several measures are used to evaluate the model performance on the test data, including the mean absolute error, 
\begin{equation}
    \textrm{MAE} = \frac{1}{M}\sum\limits_{i=1}^M\left|\Delta U_{i}^+-\Delta \tilde U_{i}^+\right |,
    \label{eq:MAE}
\end{equation}
and the mean absolute percentage error, 
\begin{equation}
 \textrm{MAPE} = \frac{1}{M}\sum\limits_{i=1}^M\left |\frac{\Delta U_{i}^+- \Delta \tilde U_{i}^+}{\Delta U_{i}^+}\right|\times100.   
 \label{eq:MAPE}
\end{equation}
Here, $M$ is the number of samples in the test data set, $\Delta U_{i}^+$ is the reference drag value obtained from DNS, and $\Delta \tilde U_{i}^+$ is the drag prediction obtained from the regression model \eqref{eq:model}. The above measures provide the absolute and relative accuracy of the regression model. The goodness-of-fit $R^2$ measure is also reported.

\subsection{Linear regression}
\label{ssec:lr_svr}
\begin{figure}
    \centering
    \includegraphics[width=.45\textwidth]{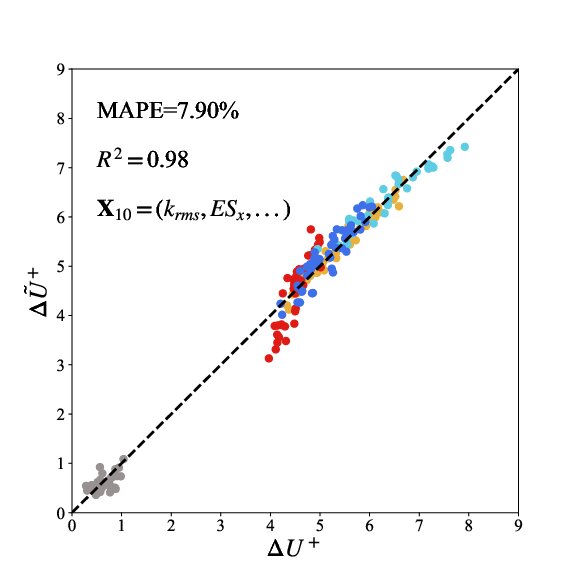}
    \includegraphics[width=.45\textwidth]{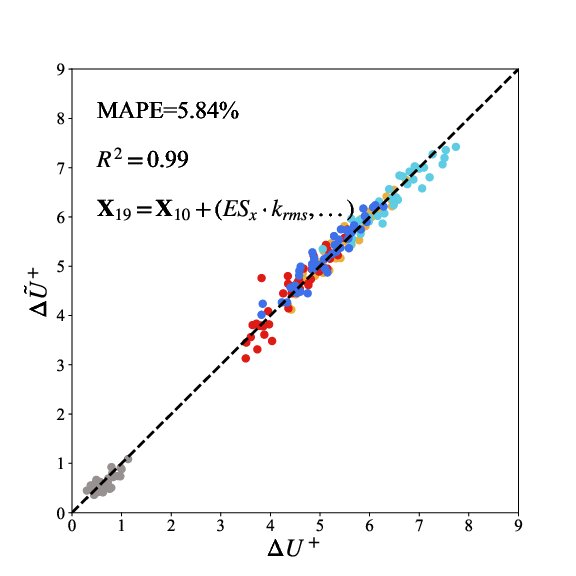}
    \put(-420, 180){(a)}
    \put(-205, 180){(b)}
    \caption{$\Delta U^+$ predictions of LR versus those from DNS: model using (a) 10 primary statistics and (b) 19 statistics (i.e. including 9 pair-product parameters.)}
    \label{fig:lr_err}
\end{figure}
We begin with the linear regression model, which is the simplest of all models considered in this study. Such a model accounts for the linear correlation between the surface parameters and $\Delta U^+$, which were observed in figure \ref{fig:corr_mat_map}. 
The model is defined as
\begin{equation}
\Delta \tilde U^+(\mathbf{x,\mathbf{w}})=\mathbf{w}^T\mathbf{x}+b.
\label{eq:LR}
\end{equation}
The weights $\mathbf{w}\in\mathbb{R}^{D\times 1}$ and the bias term $b$ are found through a least-squares optimization of the model using the training data set,
\begin{equation}
E(\mathbf{w)}=\frac{1}{2}\sum_{i=1}^N\left (\Delta U_{i}^+ - \Delta \tilde U_{i}^+\right )^2.
\label{eq:LR:objective}
\end{equation}
Figure \ref{fig:lr_err}(a) shows the drag prediction using LR on the test data samples. Using the ten primary surface-derived parameters (see table \ref{tab:stats}), the model has a MAPE of $=7.9\%$. Figure \ref{fig:lr_err}(b) shows the drag prediction obtained when using an extended number of input parameters that includes both primary and pair parameters. The extended-input model reduces the error by $2\%$, along with a decrease in data scatter (improved $R^2$). By including the pair parameters of roughness in the model input, we are incorporating non-linear effects in the linear regression model. However, the choice of pair parameters in table \ref{tab:stats} is arbitrary and we have chosen them similar to those of \cite{Jouybari:2021}. 

\subsection{Support vector regression (SVR)}
To increase the fidelity of the model, we now turn our attention to SVR, which allow for nonlinear regression through the use of kernel functions. Replacing the input vector $\mathbf{x}$ in \eqref{eq:LR} with a non-linear mapping $\bar \phi(\mathbf{x})$, we will have
\begin{equation}
    \Delta \tilde U^+(\mathbf{x})=\mathbf{w}^T\bar \phi(\mathbf{x})+b.
    \label{eq:SVR}
\end{equation}
When using kernel functions, the weight vector $\mathbf{w}$ is given by a linear combination of the expansion basis, 
\begin{equation}
    \mathbf{w}=\sum_{i=1}^N a_i \bar \phi(\mathbf{x}_i).
    \label{eq:support_vector_expansion}
\end{equation}
Inserting \eqref{eq:support_vector_expansion} into \eqref{eq:SVR} results in 
\begin{equation}
     \Delta \tilde U^+(\mathbf{x})=
     \sum_{i=1}^N a_i 
      \bar \phi(\mathbf{x}_i)^T\bar \phi(\mathbf{x})
     =  \sum_{i=1}^N a_i 
    k(\mathbf{x}_i, \mathbf{x}) + b,
    \label{eq:kernel}
\end{equation}
where ${k}(\mathbf{x}_{i}, \mathbf{x})$ is the kernel. 

In the model above, the prediction requires $N$ function evaluations. Since, $N\in \mathcal{O}(10^3)$ is the number of training samples, kernel evaluations become inefficient for large datasets. Support vector regression sparsifies the kernel by including only support vectors in the expansion. To achieve this, instead of a least-squares minimization \eqref{eq:LR:objective}, one minimizes the $\epsilon$-sensitive cost function, defined as
\begin{equation}
    J(\Delta U^+-\Delta \tilde U^+)=
    \left\{ \begin{array}{lc}
    \vspace{0.5cm}
    \left |\Delta \tilde U^+-\Delta U^+\right |-\epsilon & \quad \textrm{for} \quad \left |\Delta \tilde U^+-\Delta U^+\right |>\epsilon\\
    0 &\textrm{otherwise.} 
    \end{array} \right .
    \label{eq:SVR:objective}
\end{equation}
This means that only errors larger than $\epsilon$ contribute to the cost function. 

To determine the weights and the bias in \eqref{eq:SVR}, we minimize the regularized cost function,
\begin{equation}
E(\mathbf{w})=   C \sum_{j=1}^N J(\Delta U^+_{i}-\Delta \tilde U^+_{i})+ \frac{1}{2}\|\mathbf{w}\|^2,
\end{equation}
The second term is the regularization term that penalizes large weights, i.e. promoting flatness. Note that, by convention, the regularization parameter $C$ appears in front of the first term. The key aspect of SVR is that by using \eqref{eq:SVR:objective}, $a_j$ in \eqref{eq:kernel} are non-zero only for the training samples either lying on or above the boundary defined by $\epsilon$. 

The choice of kernel in this work for non-linear mapping is the radial basis function (RBF), 
\begin{equation}
    k(\mathbf{x}_i,\mathbf{x})=\text{exp}(-\gamma||\mathbf{x}_i-\mathbf{x}||^2),
    \label{eq:RBF}
\end{equation}
where $\gamma=1/(N\sigma^2)$ is the kernel coefficient and $\sigma^2$ is the variance of the training data. The input data $\mathbf{x}$ is rescaled by the min-max normalization while the scaling of the target $\Delta U^+$ is insignificant for prediction. The parameter $C$ and kernel bandwidth $\varepsilon$ were tuned and the best performance was obtained for values of $C=0.1$ and $\epsilon=0.01$. We have presented a simplified formulation of the optimization problem associated with SVR here. We refer to \cite{Vapnik:1995} and \cite{Smola:2002} for the complete formulation of the kernel in the optimization process, including the use of slack variables.

\subsection{Neural networks}
While SVR has far greater capacity and fidelity than LR --due to mapping the input space onto a higher-dimensional space-- it still requires the user to choose an appropriate expansion basis $\bar \phi(\mathbf{x})$. Neural networks can learn $\bar \phi$ from a broad class of functions and form a composition of such functions using hidden layers. Neural networks often require more training data than SVR to generalize well and constitute a non-convex optimization problem. To explore neural networks for drag prediction, we consider multilayer perceptron (MLP) and convolutional neural network (CNN).

\subsubsection{Multilayer perceptron (MLP)}
The MLP model is composed of multiple layers of neurons, where the neurons of two adjacent layers are connected by weights. The inputs are either the 10 primary statistics or the extended set of 19 statistics listed in table \ref{tab:stats}. The output, $\Delta \tilde U^+_{i}$, is composed from the non-linear transfer functions of each layer. This is what enables an MLP to account for high degrees of non-linearity. The objective is to identify the weights of a network such that the following loss function is minimized,
\begin{equation}
E(\mathbf{w})= \frac{1}{2}\sum_{i=1}^N \|\Delta U_{i}^+-\Delta \tilde U_{i}^+\|^2 + \frac{\lambda}{2}\mathbf{w}^T\mathbf{w}
    \label{eq:loss}
\end{equation}
The loss function is composed of a sum of squared errors term and a regularization term. The weight vector $\mathbf{w}$ contains values between the neurons of adjacent layers of the network.

We performed Bayesian optimization to determine the hyperparameters (HPs) of the MLP, including the number of layers, the number of neurons, the learning rate, the regularization term $\lambda$, the activation function, and the initialization of the weights. The Gaussian process acts as a surrogate model to estimate the model performance and the HPs are updated after each evaluation of the loss function. The acquisition function directs the next search location in the given range of parameter space to find the optimal set of HPs. At each iteration, these HPs are evaluated by training the neural network, where the number of evaluations depends on the input dimension. 

\begin{figure}
    \centering
    \includegraphics[width=.35\textwidth, angle=90]{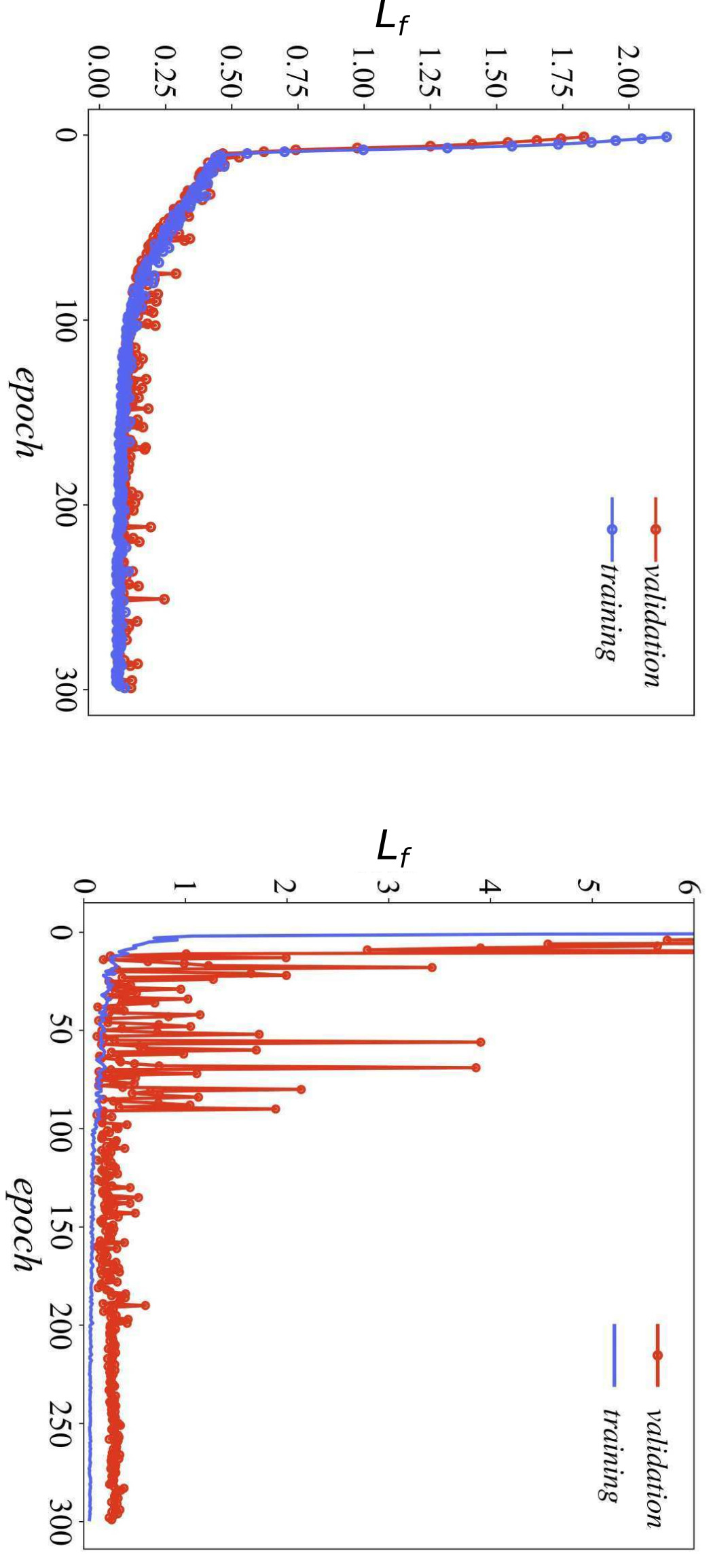}
    \put(-353, 154){\small(a)}
    \put(-172, 154){\small(b)}
    \caption{Loss curves of training and validation in the Bayesian-optimized (a) MLP$_{10}$ (b) CNN with leaning rate reschedule. Early stopping was employed during the NN within the BO process to mitigate overfitting and expedite training. }
    \label{fig:loss}
\end{figure}

Using Bayesian optimization, we developed two architectures. The first one maintains a fixed number of layers with an optimized number of neurons. The second architecture has an optimized number of layers but a fixed number of neurons. Given that each layer learns different information from the previous input, the number of neurons or filters, in theory, should differ at each layer. After conducting a set of comparative trials for both architectures, we adopted the first architecture since it exhibited a slightly lower relative error. The final hyperparameters for the two MLP models are displayed in table \ref{tab:hps_NNs}. Note that to ensure consistent scaling, the inputs were rescaled by their respective standard deviations. 

Figure \ref{fig:loss}(a) shows the training and validation losses for the MLP as a function of the number of epochs (i.e. iterations in the BO optimization process). The rapid decay of the training curve to a plateau after 100 epochs indicates a fast convergence. The validation curve --which represents the loss on a separate dataset not used for training-- follows a similar initial decay followed by a plateau. This indicates that the model generalizes to unseen data relatively well. 

\begin{table}
    \small
    \centering
    \setlength\tabcolsep{1.9pt}
    \begin{tabular}{ ccccccccccc }
    \hline
    Model & {\thead{Number of \\layers/blocks}} & {\thead{Number of \\neurons/filters}} & {\thead{Filter \\sizes}} & {\thead{Learning \\rate}} & $\lambda_2$ & {\thead{Batch \\sizes}} & {\thead{Activation \\function}} & Initialization \\ \hline 
    $\mathrm{MLP}_{10}$       & $3$    & ($256,109,256$) & N/A & $6 \times 10^{-3}$ & $2.2 \times 10^{-4}$ & ($3,2$) & leaky ReLU & Glorot uniform\\
    \vspace{1pt}
    $\mathrm{MLP}_{19}$       & $3$    & ($178,239,143$) & N/A & $1.3 \times 10^{-4}$ & $1.3 \times 10^{-4}$ & ($3,9$) & leaky ReLU & Glorot uniform\\
    \vspace{1pt}
    $\mathrm{CNN}$            & $5$    & ($64,37,64,44,30$) &($3,6,7,8,3$)& $7 \times 10^{-5}$ & $1 \times 10^{-5}$ & ($17,16$) & leaky ReLU & Glorot uniform\\
    \hline
    \end{tabular}
  \vspace{2pt}
  \captionsetup{width=0.98\textwidth,justification=justified}
  \captionof{table}{The Bayesian-optimized hyperparameters in MLP$_{10}$, MLP$_{19}$ and CNN that are used for prediction in this work.}\label{tab:hps_NNs}
\end{table}

\subsubsection{Convolutional neural network (CNN)}
This regression model is a network with convolutional layers, i.e. a set of filters (or kernels) that are convoluted with the layer's input data. One key feature is that it has sparse connectivity between the neurons, allowing for the processing of very high-dimensional input data. In our case, the input is a 2D function representing the height of the surface roughness. The objective of the CNN is to identify weights to minimize the loss function \eqref{eq:loss}.  We followed the same procedure used for the MLP to determine the architecture, i.e. the hyperparameters were obtained using Bayesian optimization. The number of blocks, filters, kernel size, learning rate, activation function and weight initialization of the CNN are reported in table \ref{tab:hps_NNs}. 

Figure \ref{fig:loss}(b) shows the training and validation losses for the CNN model. While the loss function of the training demonstrates a fast convergence, the corresponding validation curve shows large oscillations for the first 100 epochs, indicating overfitting of the unseen data. To improve CNN convergence, we implemented a learning rate schedule that reduces the rate by 0.1 every 100 epochs starting from epoch 100. Note that many other architectures are potentially more suitable for drag prediction. Our choice is, however, sufficient for comparative purposes. 

\subsubsection{Sensitivity analysis of training size}
\begin{figure}
    \centering
    \begin{minipage}{.4\linewidth}
        \centering
        \includegraphics[width=\linewidth]
        {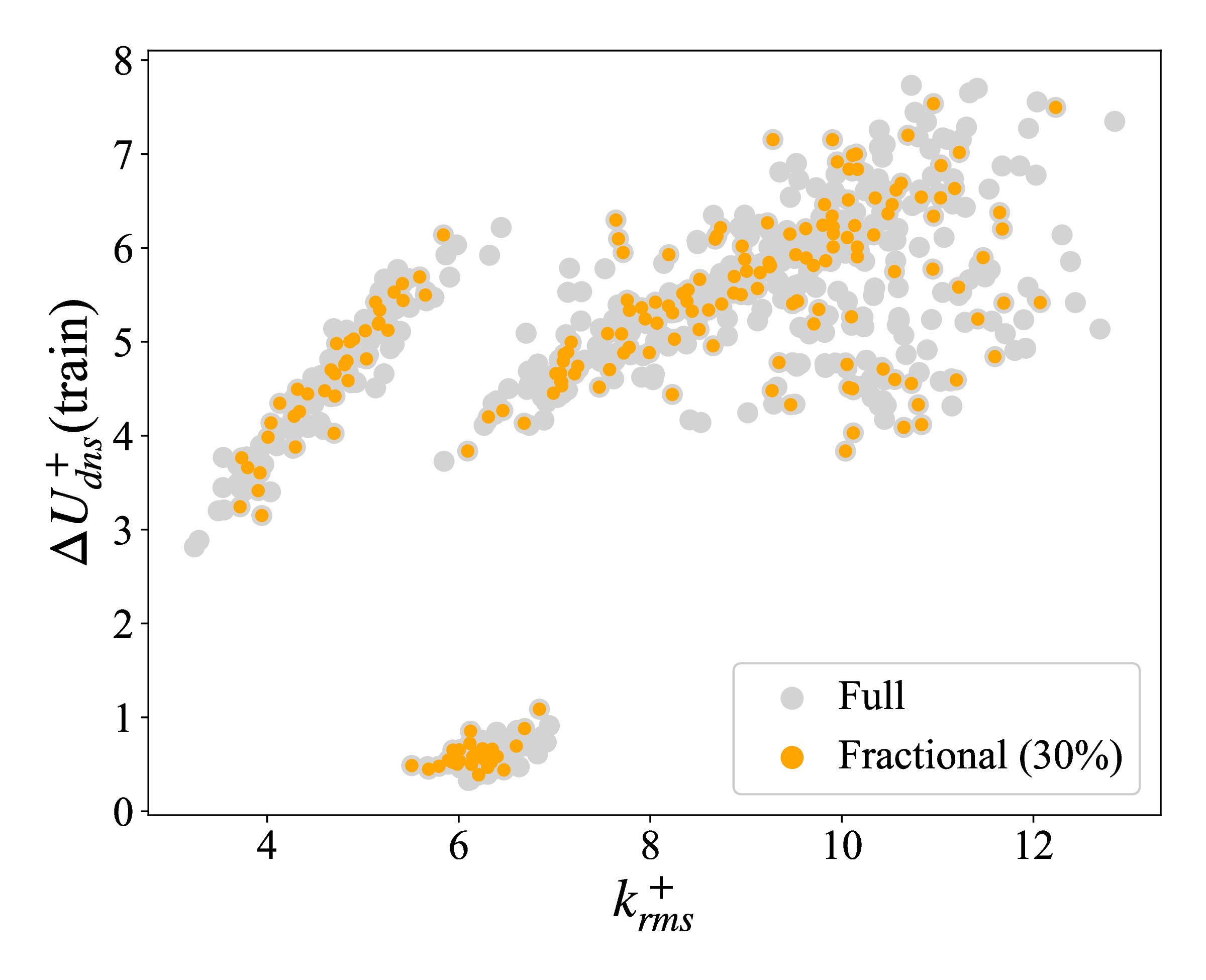}
        \put(-184, 140){(a)}
    \end{minipage}
    \begin{minipage}{.4\linewidth}
    \centering
        \includegraphics[width=\linewidth]{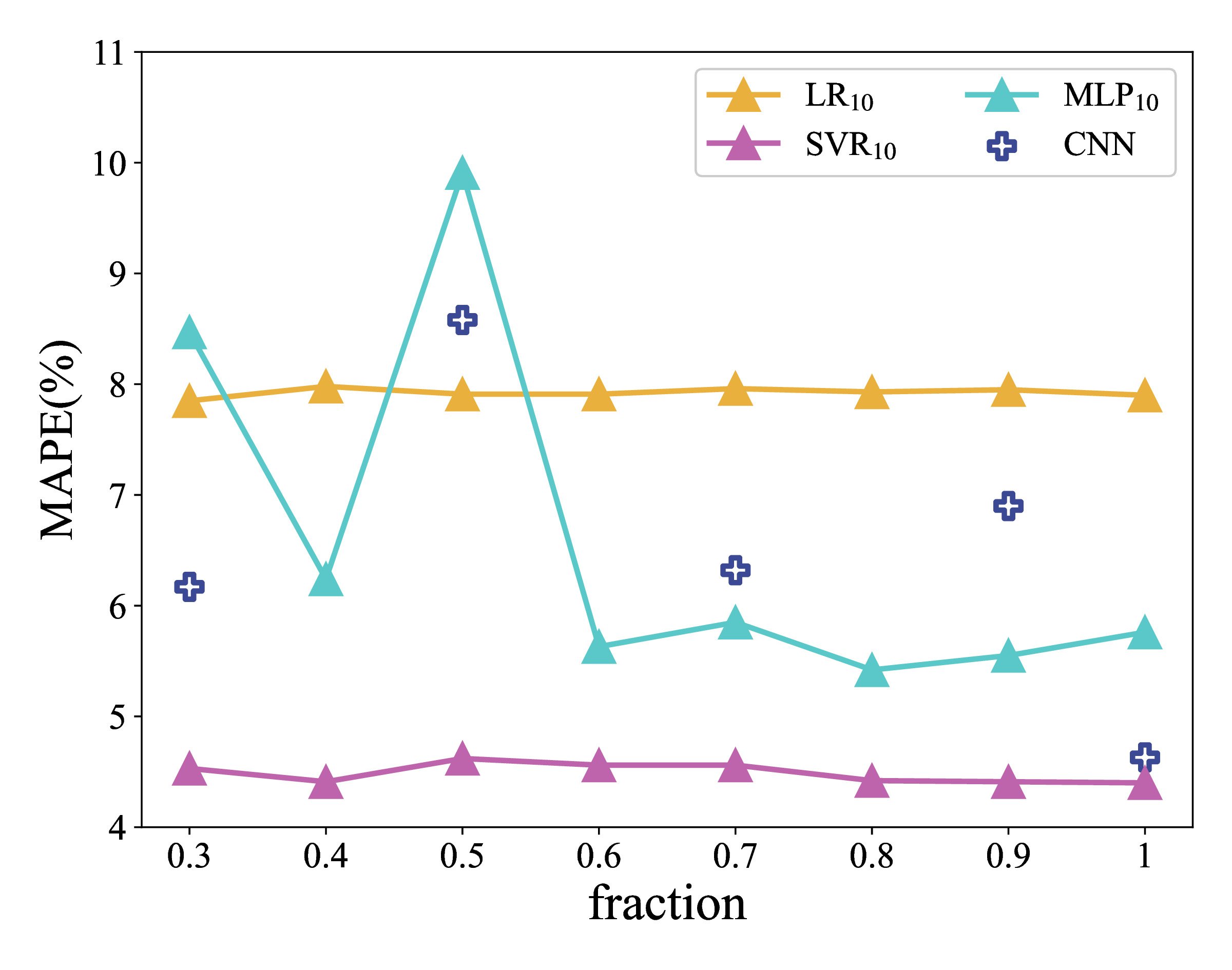}
        \put(-186, 135){(b)}
    \end{minipage}
    \begin{minipage}{0.4\linewidth}
    \centering
        \includegraphics[width=\linewidth]{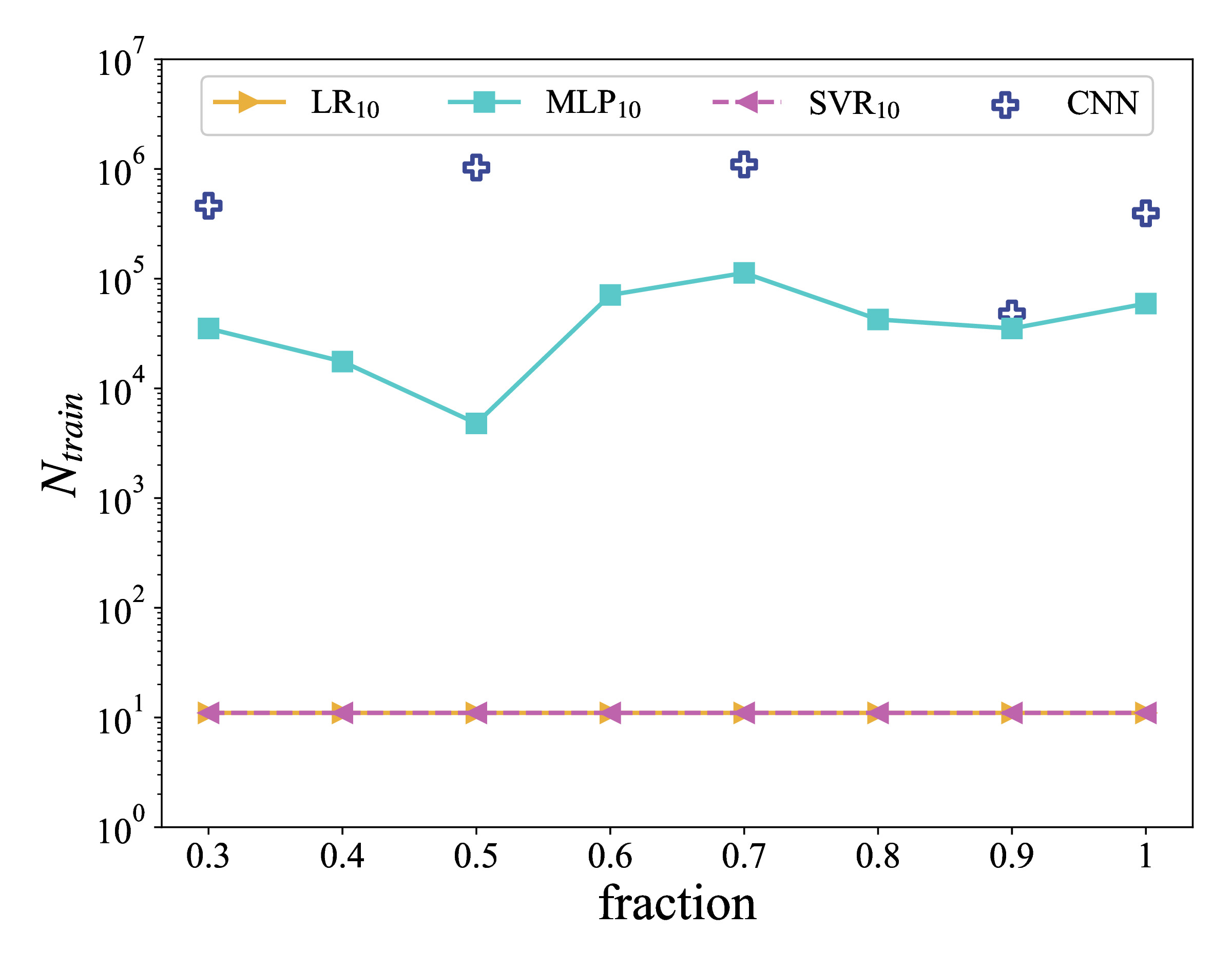}
        \put(-185, 135){(c)}
    \end{minipage}
    \caption{(a) The sample coverage in $\Delta U^+-k^+_{rms}$ space at the fraction of 30\%. The reduced training samples consistently cover the full parameter space. (b) MAPE of inference obtained from LR$_{10}$, MLP$_{10}$, SVR$_{10}$, and CNN at different sample fractions. (c) The variation in the number of trainable parameters in each model at different sample fractions.  }
    \label{fig:data_size_test}
\end{figure}

The size of training data is critical for truly exploiting the advantages of neural networks. While our dataset with over 1000 samples is, to the best of our knowledge, the largest such collection for rough-wall turbulence, it is still relatively small compared to what is commonly used for training neural networks in other applications. Therefore, we conducted a sensitivity analysis of the sample size for the training process. To ensure an even representation across different surface categories in parameter space, training samples are uniformly downscaled by the same proportion, as illustrated in figure \ref{fig:data_size_test}(a). The depth of the new neural networks (NNs) trained using varying data fractions was kept the same as the initial MLP and CNN architectures, while the number of network units were optimized using BO. 

Figure \ref{fig:data_size_test}(b) presents the relative prediction errors (MAPE) of identical test data using models trained with varying data fractions. SVR achieves the lowest error and exhibits high robustness, as its predictions remain consistent for all training data fractions. The prediction by linear regression is also not affected by the data size but consistently yields the highest error among the models compared. As expected, the performance of neural networks depends on the size of the training data. The MLP model converges with a 60\% fraction of the entire data, while the CNN model does not exhibit a clear convergence trend. Despite that, the best CNN, trained using the full dataset, achieves a low error of 4.6\%, which is comparable to SVR. Therefore, the CNN model has not yet achieved adequate generalizability to be employed for unseen data. 

Figure \ref{fig:data_size_test}(c) shows the variation of the number of trainable parameters ($N_{train}$) for the models with different training data fractions. Unlike LR and SVR, where $N_{train}$ is fixed, the NN models exhibit a non-monotonic trend. The MLP seems to stabilize around an $N_{train}$ of the order of $10^5$ after reaching a fraction of 70\%, with roughly an order of magnitude fewer trainable parameters on average than the CNN. This value is an approximation of the optimal model capacity for learning the underlying mapping. In contrast, CNN models experience a significant drop by an order of magnitude at a fraction of 90\%, followed by an increase. This behavior indicates an overfitting for CNN with the current volume of data, as was reflected by the loss in figure \ref{fig:loss}(b). The model evaluations in section \ref{sec:discussion} use predictions from models which were trained on the entire dataset.

\section{Drag prediction performance}
\label{sec:discussion}
\subsection{Comparison of regression models}
\begin{figure}
    \centering
    \includegraphics[width=.5\textwidth]{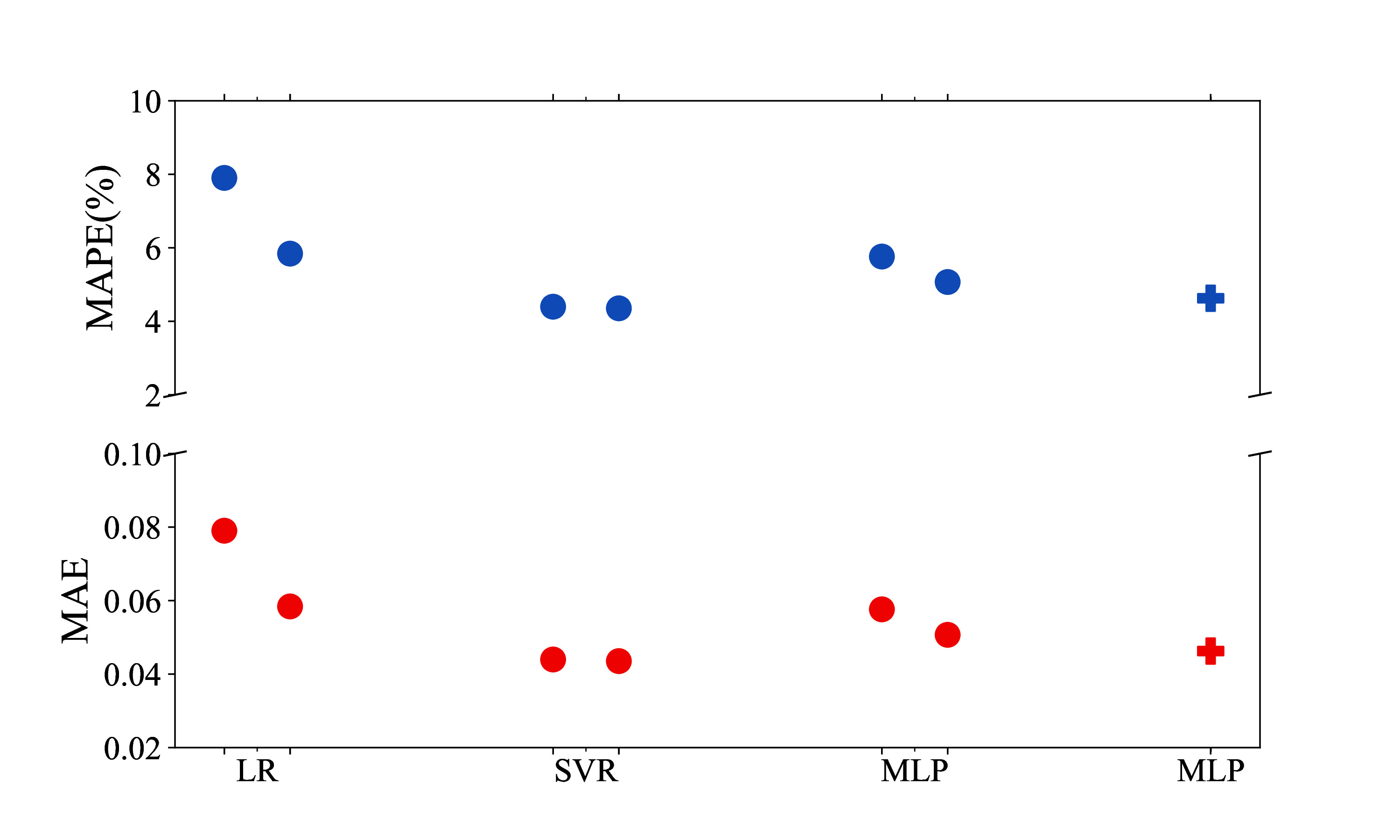}
    \put(-226, 120){(a)}
    \includegraphics[width=.5\textwidth]{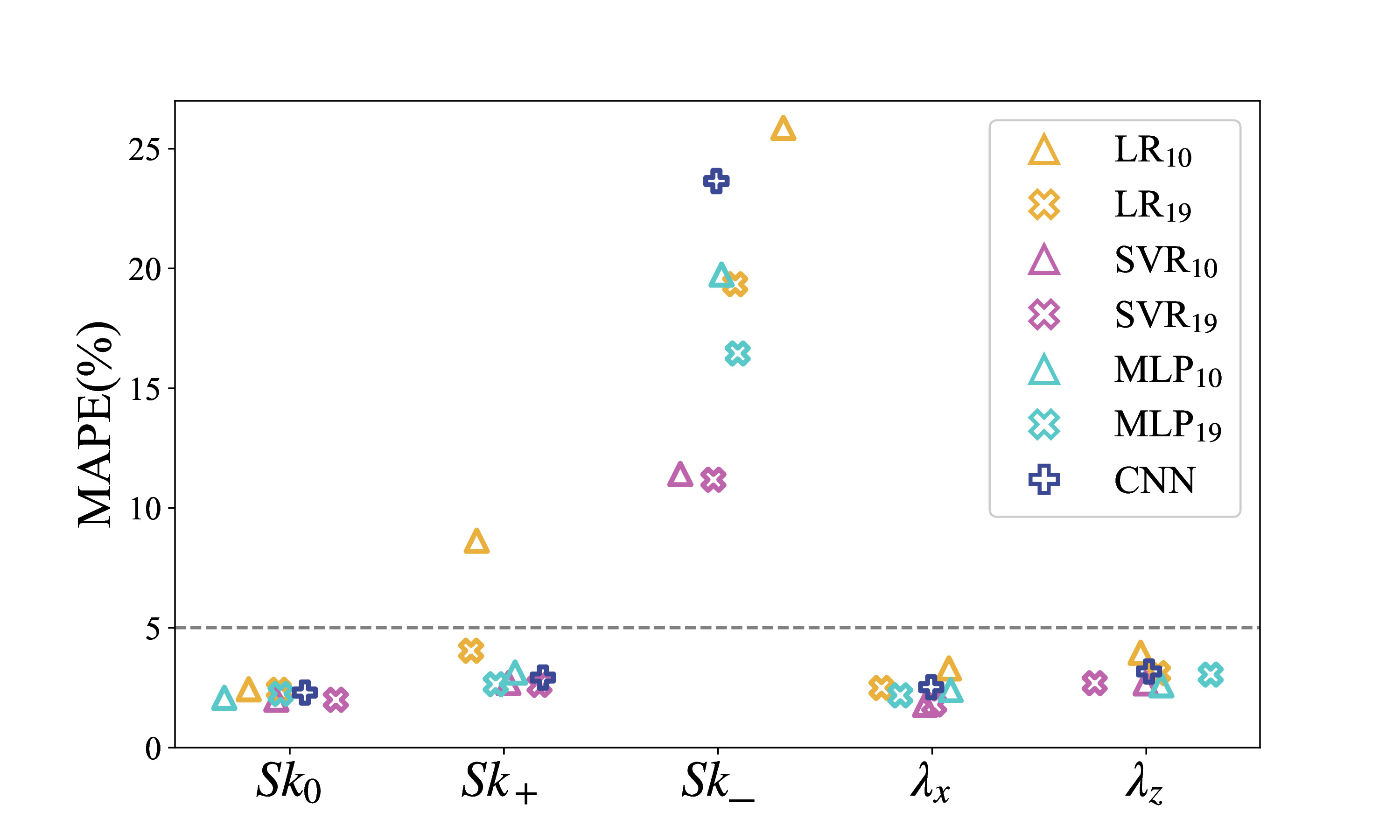} 
    \put(-226, 120){(b)} 
    \caption{(a) MAPE (\%) (blue) and MAE (red) obtained from all models trained by the hybrid data. Left and right dots correspond to the 10 and 19 parameters. All maximum errors correspond to negatively skewed surfaces (Type: $Sk_-$). (b) Applying the trained model by the full dataset on each type of surface thus the corresponding mean errors.}
    \label{fig:err_10_19}
\end{figure}

We now compare the performance of the four types of regression models. We trained LR, SVR, and MLP using two sets of inputs: ten primary statistics and 19 pair statistics as listed in table \ref{tab:stats}. We refer to these models as LR$_{10}$, LR$_{19}$, etc. The absolute error (MAE in eq.~\ref{eq:MAE}) and relative error (MAPE in eq.~\ref{eq:MAPE}) obtained from the seven evaluated models trained on the entire roughness dataset is shown in figure \ref{fig:err_10_19}(a). The LR trained using ten primary statistics displays the largest error in predicting $\Delta U^+$. As previously observed in figure \ref{fig:lr_err}, by incorporating the nine additional pair parameters, some degree of non-linearity becomes incorporated into the LR model and its error becomes reduced.

SVR emerges as the optimal predictive model with an error of 4.4\%, the smallest of all models. The use of the extended input (SVR$_{19}$), does not improve prediction compared to the SVR$_{10}$ model. This suggests that the chosen kernel function effectively captures the non-linearity embedded within the input space. Moving on to the performance of MLP, we make two observations. First, MLP$_{19}$ yields a MAE of around 5\%, which is slightly larger than the SVR model. Second, the input size has a small influence on the prediction performance, with MLP$_{19}$ performing slightly better than  MLP$_{10}$. Although the network has a near-optimal performance, these observations imply that it has not fully captured the non-linearity in the mapping from the inputs to the output. Presumably, a different network and/or larger database are required to reach the same level of performance as the SVR model.

Finally, we observe that the best-performing CNN achieves comparable results to SVR. However, it requires significantly longer training time due to its vast number of trainable parameters ($O(10^5)$). As mentioned in the previous section, to develop a generalizable CNN model a larger dataset is required. The local spatial topographical information in this specific case does not offer discernible advantage for solely predicting a single scalar value ($\Delta U^+$). However, CNN is inherently capable of learning hierarchical representations from grid-like data, which could potentially be advantageous when considering patchy, inhomogenously distributed roughness where a statistical parameterization becomes non-trivial.

\subsection{Model performance for different surface categories}
\begin{enumerate}
    \item In addition to evaluating the prediction accuracy using test data from the full roughness database, further insight into the models is gained by assessing how accurately they predict different roughness types. As shown in figure \ref{fig:err_10_19}(b), all predictive models demonstrate a comparable level of accuracy in predicting the Gaussian surfaces, i.e. $Sk_0$, $\lambda_x$, and $\lambda_z$. The average error of all models, including LR$_{10}$, is around 2-3\%.  The largest errors are found for the negatively skewed roughness, which have a pit-dominated topography. Note that MAPE is normalized with $\Delta U^+$, resulting in large errors for small values of $\Delta U^+$, which is the case for $Sk_{-}$. For example, applying LR$_{10}$ to the $Sk_{-}$ test data gives $\langle \Delta U^+ \rangle=0.60$ and $\langle  \Delta \tilde U^+ \rangle=0.64$, resulting in a large $\mathrm{MAPE} = 25$\%. This is despite the difference between the DNS and predicted drag being relatively small. Considering that MAPE is a sensitive measure for small target values, we can confirm that the SVR models notably outperform the neural networks, where the latter have an error of $\sim11\%$.
\end{enumerate}

In summary, we observe that SVR is consistently the most robust (figure \ref{fig:data_size_test}) and efficient (figure \ref{fig:err_10_19}) model in predicting the additional drag induced by homogeneously\nobreakdash-distributed irregular roughness. While the neural network models have comparable performance, they lack robustness when trained on our dataset consisting of $\mathcal{O}(10^3)$ samples. 

\subsection{Key features for SVR prediction}
We observed that SVR model's prediction performance did not benefit by extending the input with the pair parameters. In contrast, we have observed that the non-linearity introduced by using the extended input improves the prediction performance of both the LR and MLP models, which indicates that these models are unable to fully learn the non-linearity inherent in the data. In the following section, we attempt to provide insight into the capabilities of SVR.

\subsubsection{Choice and interpretation of kernels}
SVR uses a kernel to map a low-dimensional input vector to a high-dimensional space where the relationship between the inputs and the output (e.g.~$\Delta U^+$) can be mapped linearly. This feature allows SVR to implicitly take into account the pair parameters in table~\ref{tab:stats}, even though the actual input, $\mathbf{x}$, used in the model \eqref{eq:SVR} only contains the primary parameters. 
In Appendix \ref{app:kernel-example}, we present a simple example using the kernel
\[
k(\mathbf{x}_i,\mathbf{x})=\left(1+\mathbf{x}_i^T\mathbf{x}\right)^2
 \]
to illustrate how a non-linear relation between primary surface statistics and $\Delta U^+$ is transformed to a linear relation between an extended input vector and $\Delta U^+$. 

Leveraging kernels removes the arbitrariness that exists when manually selecting different combinations of the primary statistics. In fact, by constructing the kernel using radial basis functions, not only the pair parameters but an infinite number of products of the primary statistics are taken into account. To see how this works, consider the RBF kernel of eq. \eqref{eq:RBF} and assume that $N=1,D=1$ and $\gamma=1$, i.e.
\[
k(x_1,x) = e^{-(x_1-x)^2}
= e^{-x_1^2-x^2}\sum_{j=0}^\infty \left (x_1x\right)^j.
\]
We observe that last term is an infinite sum containing products of $x$. More generally, the Gaussian kernel can be interpreted as a measure of similarity between the input vector of a new surface, $\mathbf{x}$, and those of each surface in the training data set, $\mathbf{x}_i$. If $\mathbf{x}$ is close to $\mathbf{x}_i$ (in a Euclidean sense), then the corresponding term in the expansion \eqref{eq:SVR} is large. 

\subsubsection{Minimal SVR input space}
We now move on to identify the smallest set of primary statistics that is needed in the input vector $\mathbf{x}$ for accurate drag prediction using SVR. There is no need to consider products of primary parameters, since SVR implicitly takes into account all such non-linearities through the RBF kernel.  Previous work (see e.g. \cite{Daniel:2021}) have highlighted that measures of height, effective slope and skewness are necessary to capture the drag increase from homogeneously-distributed roughness. Indeed, it is likely that inputs in $SVR_{10}$ and $SVR_{19}$ contain redundancy since, for example, the first three parameters ($k^+_c$, $k^+_{rms}$, $R^+_a$) all represent the roughness height. The high correlations between these features, as seen in the triangle area demarcated in figure \ref{fig:corr_mat_map}, further support this notion.  

\begin{figure}
    \centering
    \includegraphics[width=.8\textwidth]{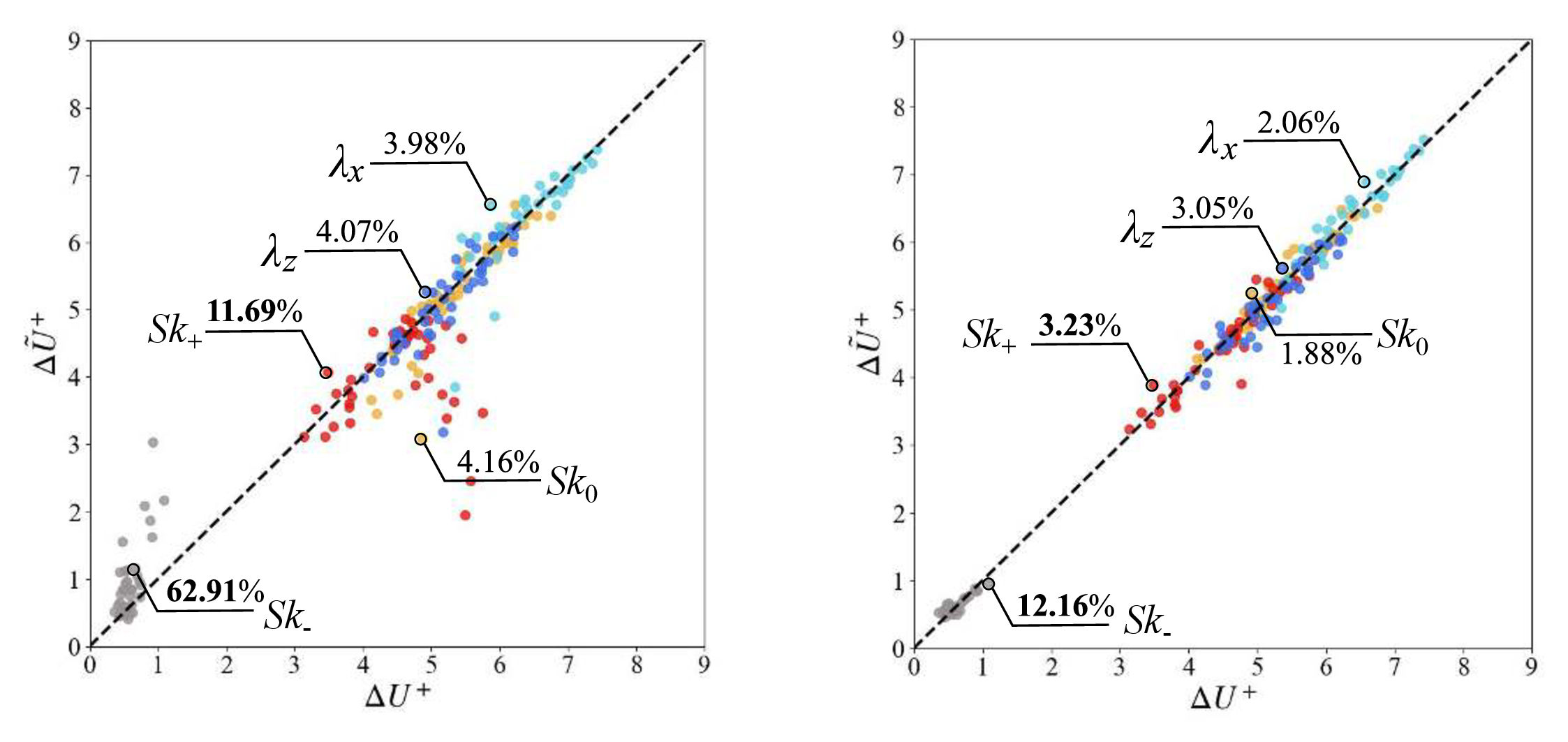}
    \put(-372, 162){(a)}
    \put(-174, 162){(b)}
    \caption{The scatter distribution of $\Delta U^+$ .vs. $\Delta \tilde U^+$ obtained from the new SVR. The model is trained with the reduced input space involving (a) $k^+_{rms}$, $ES_x$, $ES_z$ (b) and additional $Skw$. The error reduction is observed for $Sk_-$- and $Sk_+$-roughness, marked in bold. }
    \label{fig:svr_stats4}
\end{figure}

We first focus on the case where the input is $\mathbf{x}=(k^+_{rms}, ES_x, ES_z)$ and is comprised of only the vectors of the three parameters. A new SVR model, $SVR_3$, is trained on all types of roughness using these three parameters, followed by inference on the same testing data employed earlier. Figure \ref{fig:svr_stats4}(a) shows the diagonal spread of $\Delta U^+$ (obtained from DNS) and $\Delta \tilde U^+$ (predicted) for each category. A reasonably good agreement for Gaussian surfaces ($Sk_0$, $\lambda_x$, and $\lambda_z$) is observed with a mean error of $\sim4\%$. However, for $Sk_+$ surfaces, the prediction exhibits notable deviations from the DNS data, which cause the mean error to become $\sim12\%$, in contrast to the significantly lower errors of $\sim2.7\%$ obtained from SVR$_{10}$ and SVR$_{19}$. Such outliers are also observed for $Sk_-$. 
Furthermore, we tried to replace $k^+_{rms}$ with $k^+_c$ and $R^+_a$ for SVR training. While using the crest height, $k^+_c$, still gave a model with reasonable prediction accuracy, it was still inferior compared to models which used $k^+_{rms}$ and $R^+_a$. This finding suggests that $k^+_{rms}$ and $R^+_a$ provide a more comprehensive quantification of the surface terrain than $k^+_c$. 

Measures of roughness height and effective slopes, or any non-linear combination of them, are apparently insufficient to represent the skewed surfaces. We therefore trained model $SVR_4$ where the input was $\mathbf{x}=(k^+_{rms}, ES_x, ES_z, Skw)$. Figure \ref{fig:svr_stats4}(b) shows a significant error reduction for $Sk_-$ and $Sk_+$, demonstrating the essential role in rectifying all deviant samples for non-Gaussian surfaces. This result aligns with the observations of previous studies showing that skewness plays an important role in drag prediction \citep{Flack:2016, Frooghi:2017, Marchis:2020, Busse:2023}. 

\subsection{Model performance using minimal input space}
\begin{figure}
    \centering
    \includegraphics[width=.55\textwidth]{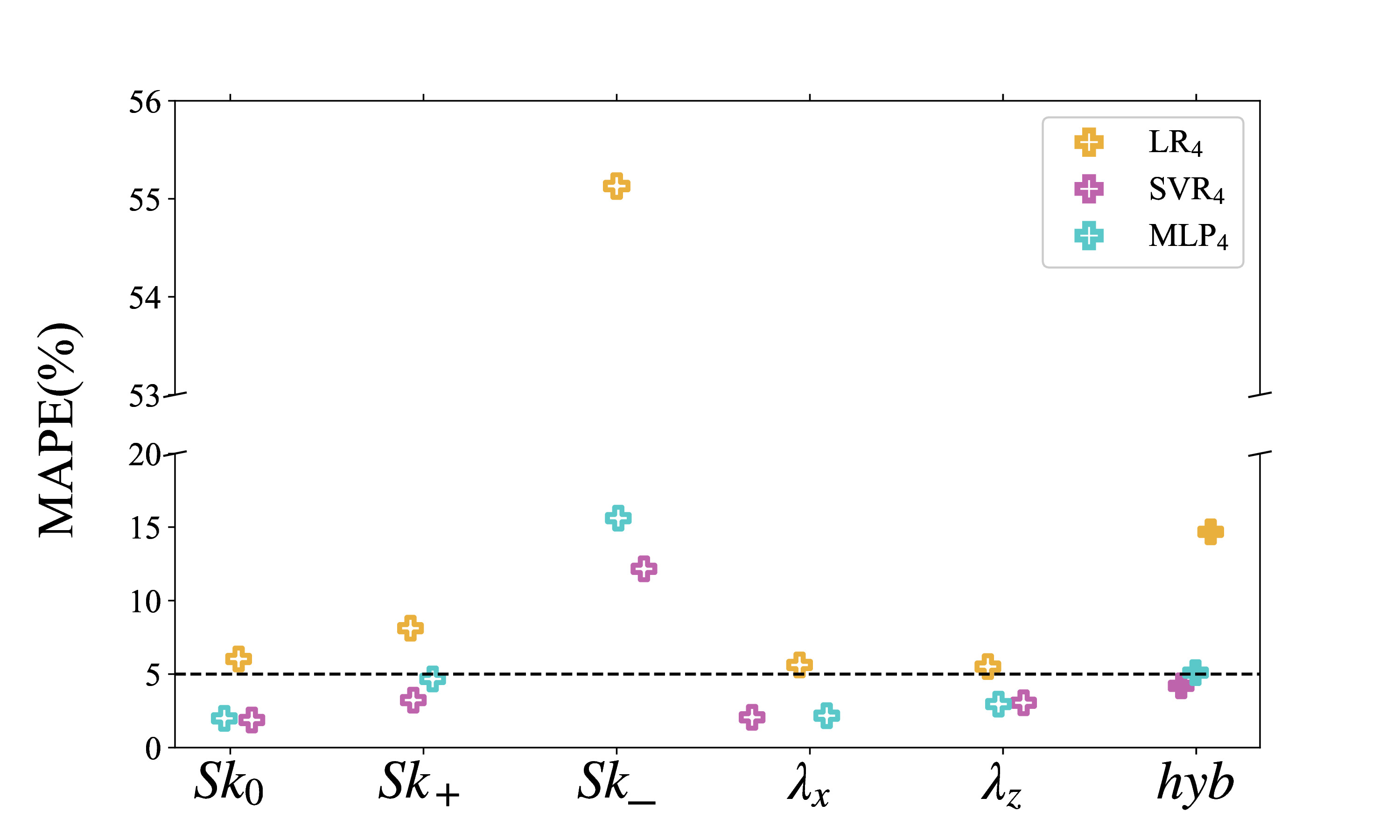}
    \caption{The average errors of each roughness category (empty circles) obtained from the three models (LR$_4$,SVR$_4$,MLP$_4$) trained by the reduced input $\mathbf{x}=(k^+_{rms}, ES_x, ES_z, Skw)$. The solid crosses represent the mean error of all surfaces.}
    \label{fig:err_4}
\end{figure}

Our findings suggest that skewness, in combination with $k^+_{rms}$ and effective slopes in both directions, can satisfactorily predict $\Delta U^+$. To investigate the influence of model complexity on key features, we trained LR and MLP models using only these four parameters. Figure \ref{fig:err_4} compares MAPE for each roughness type and all roughness types combined (denoted as \textit{hyb}). LR$_4$ achieves a MAPE of $\sim15\%$ for all-surface prediction, which is twice the error obtained by LR$_{10}$. In contrast, SVR$_4$ and MLP$_4$ maintain low errors of around 5\%, which is comparable to the performance of SVR$_{10}$ and MLP$_{19}$. For Gaussian surfaces, LR$_4$ can still yield a reasonably accurate prediction, while its performance deteriorates for negatively-skewed surfaces, resulting in $\mathrm{MAPE}=55\%$. 

\section{Discussion} \label{sec:conclusion}
In this work, we assessed data-driven regression models of different complexity to provide insight into the most appropriate modeling choice for rough-wall turbulence drag prediction. To achieve this, we generated $1018$ surface samples comprised of five categories of homogeneous roughness (Gaussian/non-Gaussian, isotropic/anisotropic). To train and test the models, direct numerical simulations at $Re_\tau=500$ were used to generate corresponding drag values, both in the transitionally- and fully\nobreakdash-rough regimes. The database is sufficiently large to allow for comparison of different regression models; including linear models, support vector regression, and neural networks. However, one should bear in mind that there exists many types of roughness distributions (e.g. patchy) and flow configurations (e.g. pressure gradients) that we have not considered. 

For datasets of size $\mathcal{O}(10^3)$ or smaller, kernel-based SVR serve as a very competitive tool for drag prediction --in particular when only a few statistical measures of the surface are available. We have shown how kernels transform non-linear mapping between the surface statistics ($k_{rms}^+, ES_x, ES_z, Skw, \dots)$ and $\Delta U^+$ into a linear one via a so-called feature map function $\bar \phi$. Specifically, using support vector regression with radial basis functions produces an efficient prediction model. For example, SVR predictions for the Gaussian ($Sk\approx0$) and peak-featuring ($Sk\gtrsim0$) surfaces had mean errors of $\sim 3\%$. We also demonstrated that an SVR model that uses four surface measures ($Skw$, $k_{rms}$, $ES_x$ and $ES_z$) provides nearly as good performance as one using $10$ or $19$ measures.

We demonstrated that linear regression methods can be expected to deliver a mean error of around 10\% for homogeneous rough surfaces, which may be sufficient for some applications. However, the model uncertainly is high, with errors reaching 25\% for negatively skewed surfaces falling into in the transitionally rough regime. The model captures the linear correlations in the data, which can be significant, for example, between $k_{rms}^+$ and $\Delta U^+$ for Gaussian type roughness. 
However, linear regression cannot capture nonlinear correlations in the data, for example, between $Skw$ and $\Delta U^+$. Nevertheless, a linear mapping between surface characteristics and drag accounts for a significant amount of physical information for most surface categories. 

Neural networks can be regarded as an extension of kernel\nobreakdash-based methods, because the basis functions $\bar\phi(\mathbf{x})$ (and thus the kernel) can, alongside the weights $w$, be tuned during the training. While the prediction accuracy of MLP is on par with SVR, it still depends on the input data size. This can be explained by the fact that the optimization problem for NN is not convex. SVR in contrast, is a convex optimization problem and will therefore find the global minimum of the objective function \eqref{eq:SVR:objective}. In addition, the number of trainable parameters of MLP is three orders of magnitudes larger than LR and SVR. It therefore seems that --for our dataset-- MLP does not offer any clear advantage over SVR. The true advantage of MLP emerges for much larger training datasets, allowing for the many trainable parameters to be adequately tuned. In theory, a two-layer MLP can approximate any continuous function on a compact input domain if the network has sufficiently many hidden units. SVR, on other hand, becomes rapidly inefficient for large training datasets, since the number of expansion terms in the model is of the same order of magnitude as the number of training samples.

We also investigated convolutional networks for drag prediction using the roughness height distribution of a surface as the input data. We did not design a CNN that is optimal for drag prediction, as the purpose of our study was merely to understand the advantages and limitations between different regression approaches. We found that CNN performs satisfactory, but the model does not generalize well, which is not surprising as the number of trainable parameters is very large compared to the size of the dataset. CNNs, however, have features that make them particularly interesting for inhomogeneous rough surfaces, i.e. roughness with features that vary spatially on a length scale comparable to the system scale (e.g.~pipe radius, boundary layer thickness). The reason is that mapping the full surface topography to drag can be considered as a pattern recognition problem. CNN is well-suited for such problems, allowing invariances to be built into the architecture. For example, when feeding the roughness height of a surface into a model, one may expect that the output should be independent of the exact position of a particular roughness element with respect to other elements. Such translational invariance can be built into the structure of a CNN. Another reason why CNNs are appropriate is because it is highly non-trivial to characterize inhomogenous roughness using statistical measures, motivating the direct use of surface height distribution.

A key takeaway from this study is that kernel-based SVRs possess high-enough fidelity for drag prediction, at least for the foreseeable future where very large databases would be unavailable. In time --and perhaps through a collective community effort-- it is likely that a sufficient amount of relevant roughness data will become accumulated to facilitate the development of efficient prediction models. We anticipate that different regression models, some of them studied here, will be suitable for different applications.

\appendix
\counterwithin{figure}{section}
\counterwithin{table}{section}

\section{Validation of DNS solver for turbulent channel flow over irregular roughness}\label{app:validation}

We chose the irregular rough surface of \citet*{Jelly:2019} for validation purposes which belongs to a DNS dataset made publicly available by the authors. The height map of the surface is shown in \hyperref[fig:Jelly_Busse_Surface]{figure \ref*{fig:Jelly_Busse_Surface}}.
\\
\par
\noindent
\begin{minipage}{\textwidth}
\makeatletter
\def\@captype{figure}
\makeatother
    \begin{center}
     \includegraphics[width=0.95\linewidth]{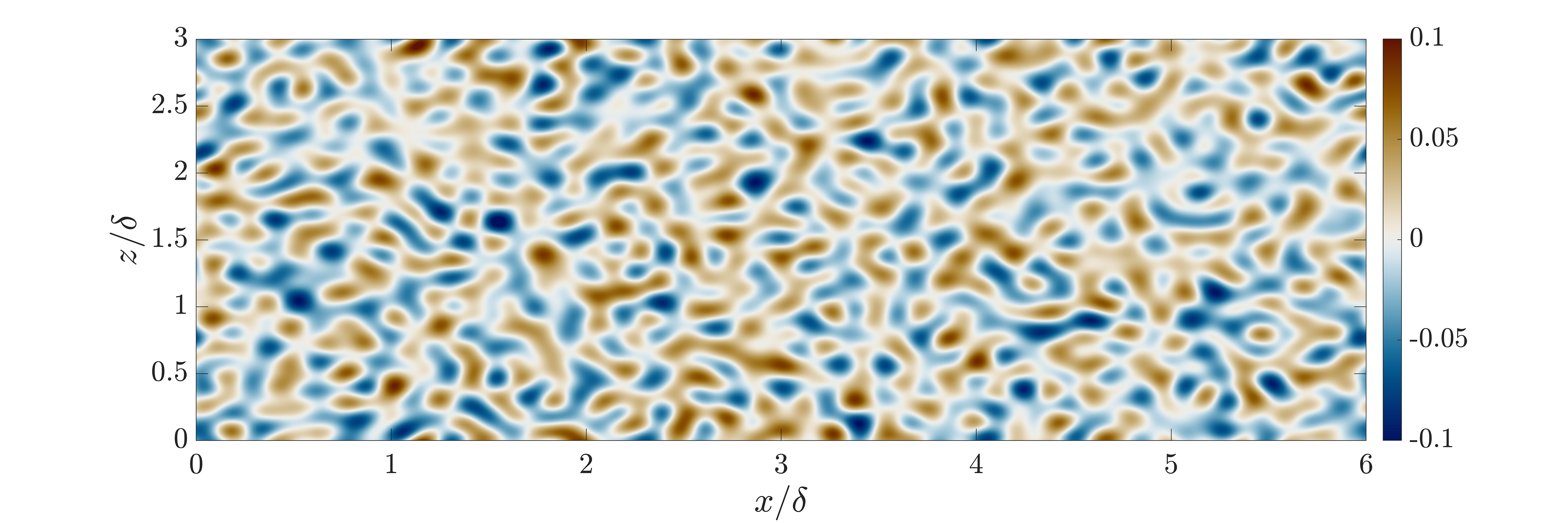}
    \captionsetup{justification=justified}
    \caption{Height map of the irregular rough surface from \cite{Jelly:2019}. Color-bar values indicate the surface height relative to the mean reference plane, i.e., $k/\delta$.}\label{fig:Jelly_Busse_Surface}
    \end{center}
\end{minipage}
\\
\par
To minimize uncertainties in trying to reproduce the results of \cite{Jelly:2019}, we endeavored to match their DNS parameters as closely as possible. Therefore, the grid resolution was chosen to be the same as theirs and the stretched wall-normal grid was also taken from their dataset. These and other relevant parameters are listed in \hyperref[tab:validation]{table \ref* {tab:validation}}.
\\
\par
\noindent
\begin{minipage}{\textwidth}
\makeatletter
\def\@captype{table}
\makeatother
  \centering
  \rule{\linewidth}{0.5pt}
  \small
  \begin{tabular}[]{ m{2.0cm}m{0.5cm}m{0.5cm}m{2.0cm}m{0.5cm}m{0.5cm}m{0.7cm}m{0.0cm} }
    $Re_{\tau}$ & ${L_x}/\delta$ & ${L_y}/\delta$ & ${L_z}/\delta$ & $\Delta x^+$ & $\Delta y^+_{\mathrm{min}}$ & $\Delta y^+_{\mathrm{max}}$ & $\Delta z^+$\\
    &&&&&&&\\
    $180$               & $6.0$ & $3.0$ & $2.0$ & $2.81$ & $0.67$ & $5.00$ & $2.81$\\
  \end{tabular}
  \captionsetup{justification=justified}
  \captionof{table}{Simulation parameters for the validation case of the irregular rough surface of \hyperref[fig:Jelly_Busse_Surface]{\ref*{fig:Jelly_Busse_Surface}}: Friction Reynolds number ($Re_{\tau}$); Domain lengths along streamwise ($L_x$), wall-normal ($L_y$) and spanwise ($L_z$) directions; Viscous-scaled streamwise grid spacing ($\Delta x^+$), minimum wall-normal grid-spacing ($\Delta y^+_{\mathrm{min}}$); maximum wall-normal grid-spacing($\Delta y^+_{\mathrm{max}}$); spanwise grid-spacing ($\Delta z^+$). Additional details and descriptions can be found in \cite{Jelly:2019}.}\label{tab:validation}
  \rule{\linewidth}{0.5pt}
\end{minipage}
\\
\par
\noindent
\begin{minipage}{\textwidth}
\makeatletter
\def\@captype{figure}
\makeatother
    \begin{center}
    \begin{subfigure}[b]{0.34\textwidth}
    	\centering
        {\captionsetup{position=bottom, labelfont=it,textfont=normalfont,singlelinecheck=false,justification=raggedright,labelformat=parens}\caption{}}
    	\includegraphics[width=\textwidth]{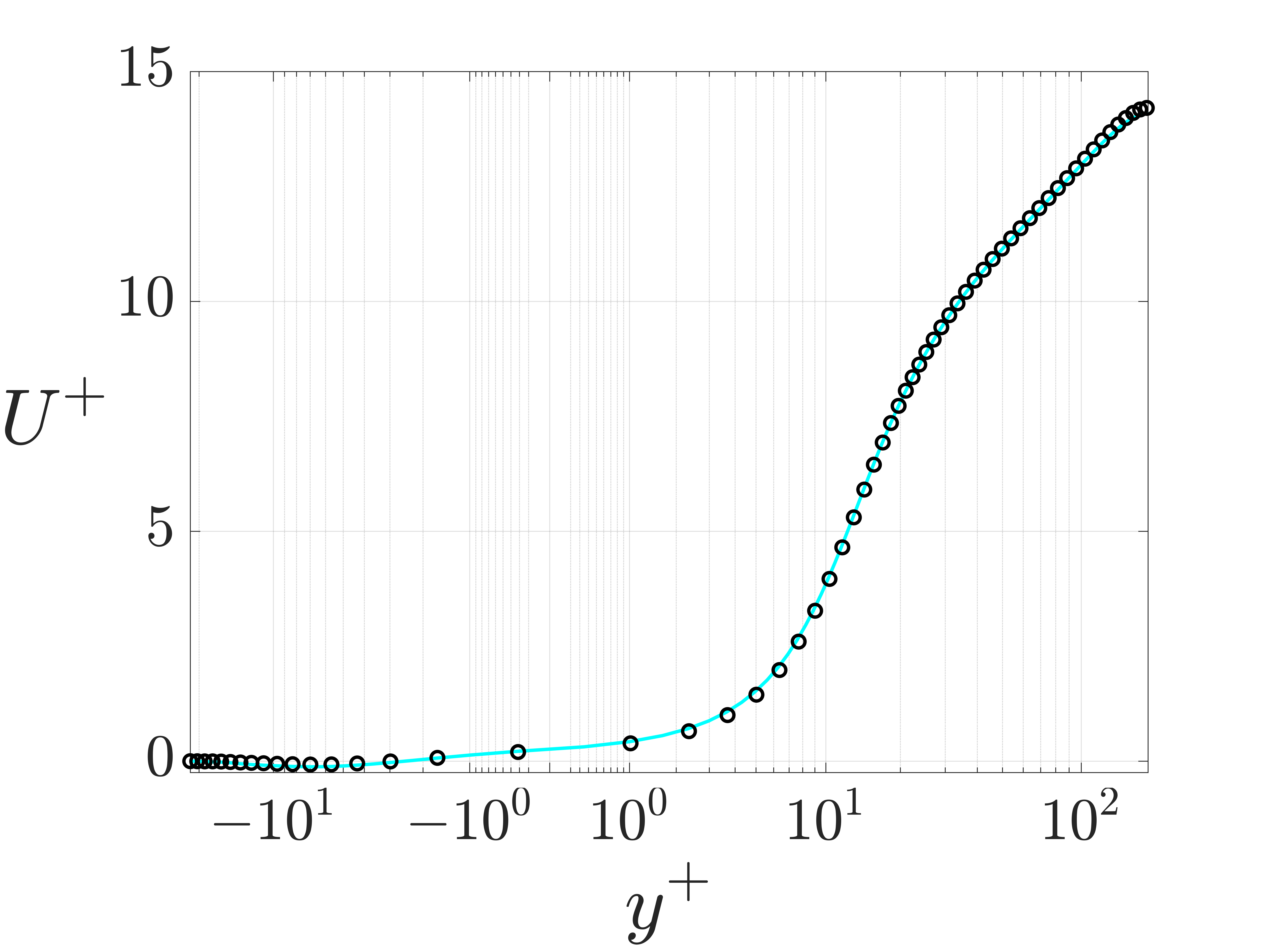}
    \end{subfigure}%
    \begin{subfigure}[b]{0.34\textwidth}
    	\centering
        {\captionsetup{position=bottom, labelfont=it,textfont=normalfont,singlelinecheck=false,justification=raggedright,labelformat=parens}\caption{}}
    	\includegraphics[width=\textwidth]{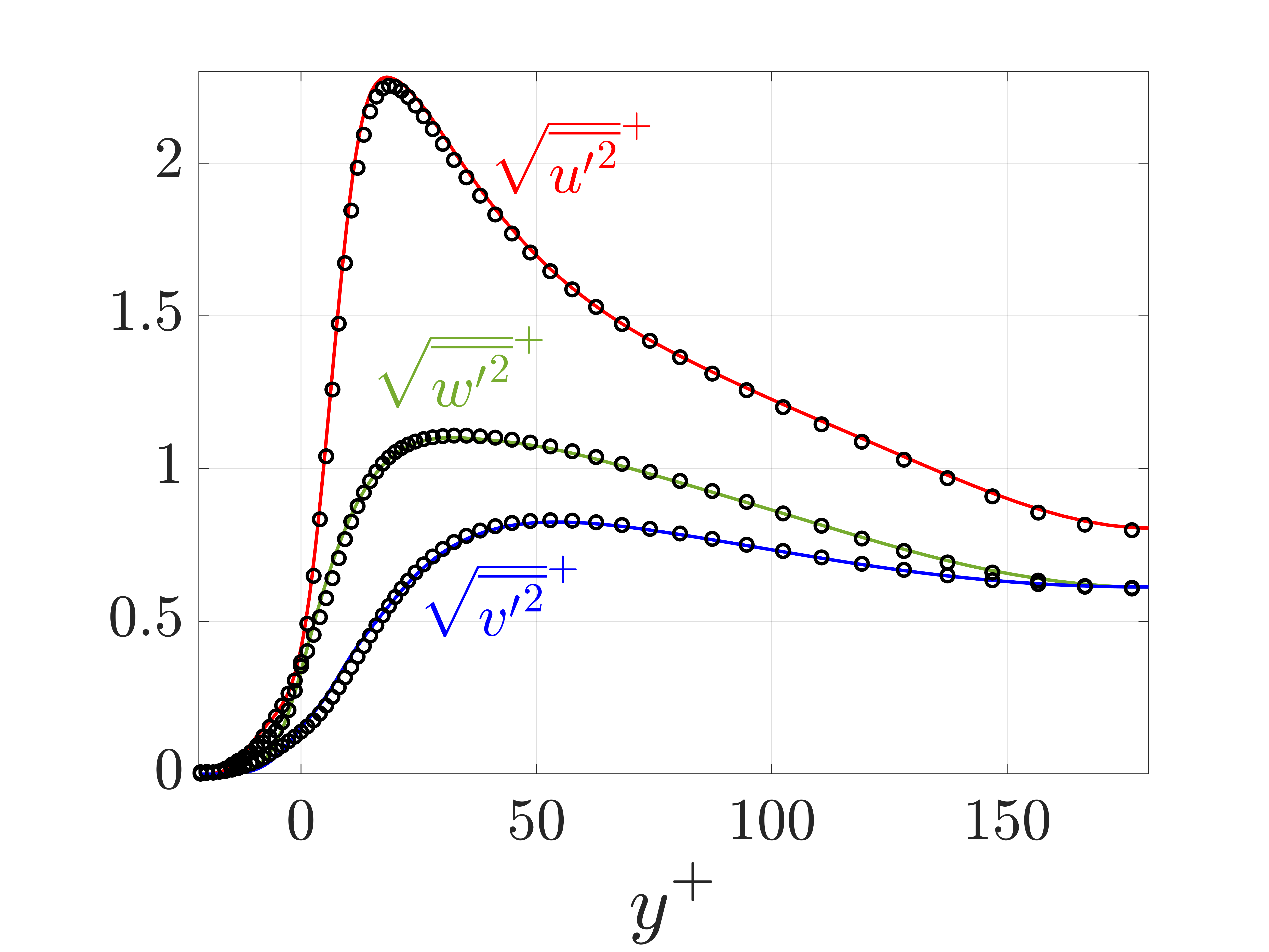}
    \end{subfigure}%
    \begin{subfigure}[b]{0.34\textwidth}
    	\centering
        {\captionsetup{position=bottom, labelfont=it,textfont=normalfont,singlelinecheck=false,justification=raggedright,labelformat=parens}\caption{}}
    	\includegraphics[width=\textwidth]{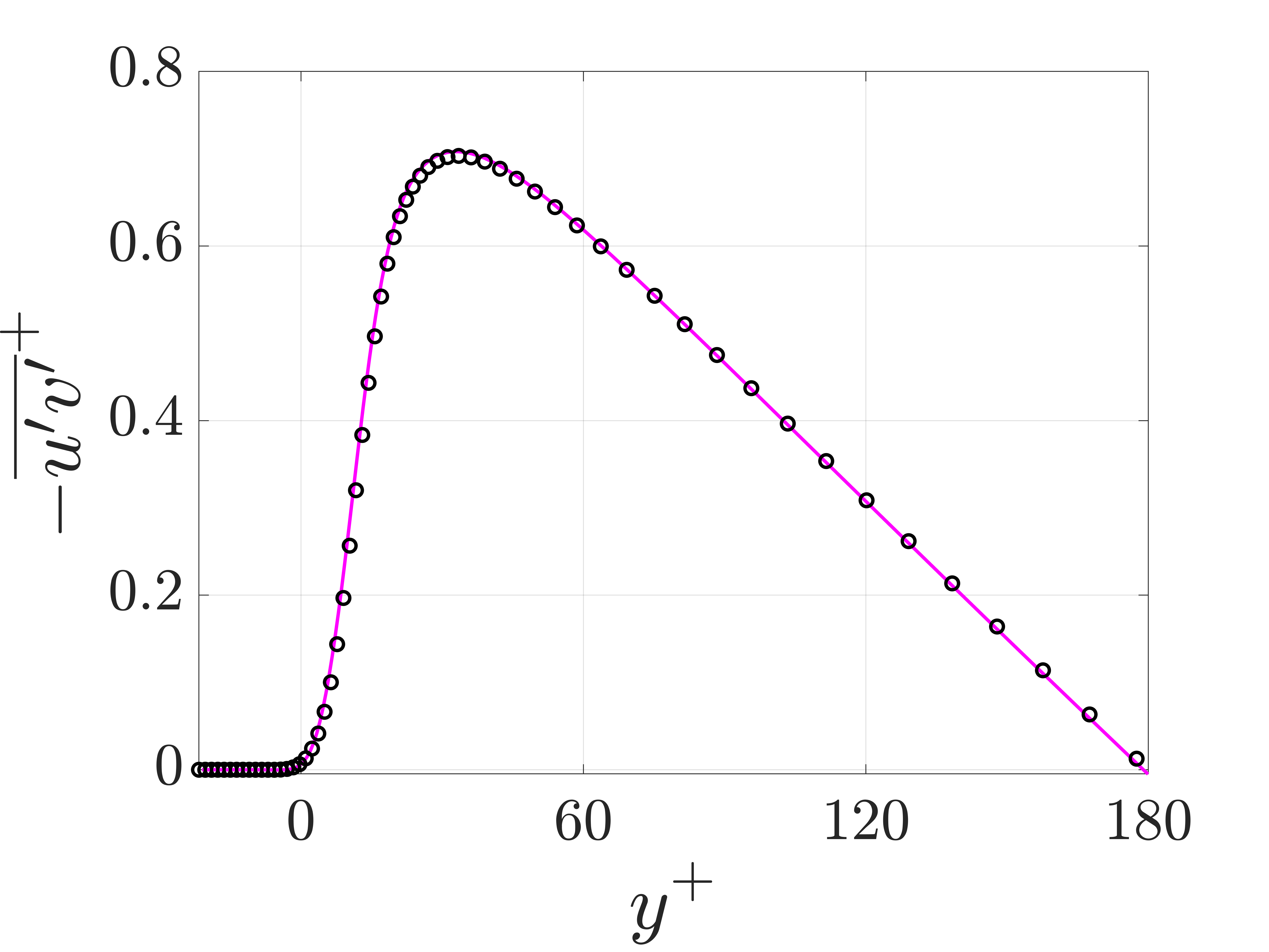}
    \end{subfigure}
    \captionsetup{justification=justified}
    \caption{Validation results: (\emph{a}) mean velocity; (\emph{b}) root-mean-square velocity fluctuations; (\emph{c}) Reynolds shear stress. Lines are the results using CaNS and symbols are the data from \cite{Jelly:2019}.}
    \label{fig:validation}
    \end{center}
\end{minipage}
\\
\par
\hyperref[fig:validation]{Figure \ref*{fig:validation}} compares the time and plane-averaged mean flow and fluctuations obtained using CaNS to those of \cite{Jelly:2019}. The agreement is good across all of the quantities considered, thereby validating the DNS methodology used to obtain the $\Delta U^+$ values for the rough surfaces of this study.

\section{Kernel methods: An example}\label{app:kernel-example}
Consider an input vector $\mathbf{x}=(x_1,x_2)\in \mathbf{R}^{2}$ 
%that is mapped to $\mathbf{R}^{6}$ using 
%\[
%\bar \phi(\mathbf{x})=(1,\sqrt{2}x_1,\sqrt{2}x_2,x_1^2,\sqrt{2}x_1x_2,x_2^2)
%:\mathbf{R}^{2}\rightarrow \mathbf{R}^{6}
%\]
and the kernel 
\[
k(\mathbf{x}_i,\mathbf{x})=\left(1+\mathbf{x}_i^T\mathbf{x}\right)^2
 \]
where $\mathbf{x}_i=(x_{i,1},x_{i,2})$ corresponds to the statistics of the $i$th surface in the training dataset of size $N$. 
The corresponding kernel-based model \eqref{eq:kernel}
%with a 2D input vector can be written as linear system with a 6D input vector
can be written as 
\begin{equation}
     \Delta \tilde U^+(\mathbf{x})
     % = \mathbf{w}^T\bar\phi(\mathbf{x)} + b = 
 %    \sum_{i=1}^N \alpha_i 
  %   \bar \phi(\mathbf{x}_i)^T\bar \phi(\mathbf{x})
     =  \sum_{i=1}^N a_i 
    \left(1+\mathbf{x}_i^T\mathbf{x}\right)^2 =
\mathbf{ w}^T\mathbf{\hat x}
    \label{eq:kernel-example}
\end{equation}
%
%with $\mathbf{x}_i=(x_{i,1},x_{i,2})$ corresponds to the statistics of the $i$th surface in the dataset of size $N$ used for training the model. Moreover, 
with $\mathbf{\hat x}=(1, \sqrt{2}x_1,\sqrt{2}x_2, x_1^2,x_2^2,\sqrt{2}x_1x_2) \in \mathbf{R}^{6}$.
Specifically, if the 2D input vector  to the kernel method is $\mathbf{x}=(ES_x,Skw)$ then there is an equivalent linear regression model with a 6D input vector $\mathbf{\hat x}=(1, ES_x,Skw, ES_x^2, Skw^2,ES_x\cdot Skw)$ in the transformed space. 
This example illustrates how the pair parameters included explicitly in the input vector of the linear regression model \eqref{eq:LR} are implicitly taken into account using a kernel method. 

Moreover, the weights $\mathbf{w}=(w_0,\dots,w_5)$ in \eqref{eq:kernel-example} are given by 
\begin{eqnarray}
    w_0&=&\sum_{i=1}^N a_i,\qquad
    w_1=\sum_{i=1}^N \sqrt{2}a_ix_{i,1},\qquad
    w_2=\sum_{i=1}^N \sqrt{2}a_ix_{i,2},\\
    w_3&=&\sum_{i=1}^N a_ix^2_{i,1},\qquad    
    w_4=\sum_{i=1}^N a_ix^2_{i,2},\qquad   
    w_5=\sum_{i=1}^N \sqrt{2}a_ix_{i,1}x_{i,2}
\end{eqnarray}
which demonstrates that they depend on linear combination of all the training data with non-zero expansion coefficients $a_i$. 
These coefficient depend on the cost function and can be obtained by using an adjoint formulation of the optimization problem. For a cost function composed of  sum of squares of error with a regularization term one can derive (see. e.g. \cite{bishop2007}) the explicit dependence of $a_i$ on the training data $\mathbf{x}_i$ and the corresponding ground truth $\Delta U_i^+$. It becomes clear that large $w_i$ indicates a high sensitivity to the corresponding input in training data set. For example, $w_3>w_2$ means that -- in the training data set -- a variation of $ES_x^2$ results in a larger change in $\Delta U^+$ compared to $Skw^2$.

\bibliographystyle{apalike}
\bibliography{references}  %%% Uncomment this line and comment out the ``thebibliography'' section below to use the external .bib file (using bibtex) .

\end{document}